\documentclass[amsmath,amssymb,nofootinbib]{revtex4}
\usepackage{dcolumn}
\usepackage{lipsum}
\usepackage{bm}
\usepackage{graphicx}
\usepackage{color}
\usepackage{amsmath}
\newcommand{\be}{\begin{equation}}
\newcommand{\ee}{\end{equation}}
\newcommand{\ba}{\begin{eqnarray}}
\newcommand{\ea}{\end{eqnarray}}

\newcommand{\n}[1]{\label{#1}}

\begin{document}

\title{Ingoing Eddington-Finkelstein Metric of an Evaporating Black Hole
\footnote{Alberta Thy 26-16, arXiv: 1607.05280 [hep-th]}}
\author{Shohreh Abdolrahimi$^{a~ b}$ }
\email{abdolrah@ualberta.ca}
\author{Don N.\ Page$^{a}$}
\email{profdonpage@gmail.com}
\author{Christos Tzounis$^{a~b}$}
\email{tzounis@ualberta.ca}
\affiliation{${}^a$Theoretical Physics Institute, University of Alberta, Edmonton, AB T6G 2E1, Canada\\
${}^b$Department of Physics and Astronomy, California State Polytechnic University, 3801 West Temple Ave., Pomona, CA 91768, USA }

\date{2019 December 12}

\begin{abstract}

We present an approximate time-dependent metric in ingoing Eddington-Finkelstein coordinates for an evaporating nonrotating black hole as a first-order perturbation of the Schwarzschild metric, using the linearized backreaction from a realistic approximation for the stress-energy tensor for the Hawking radiation in the Unruh quantum state.

\end{abstract} 

\maketitle


\section{Introduction}
The physics of black holes is an abundant field in which the convergence of gravitation, quantum theory, and thermodynamics takes place. The original derivation of Hawking radiation \cite{Hawking} from black holes is based on semiclassical effective field theory. Normally, quantum fields are considered test fields in the curved spacetime of a classical background geometry. 
A quantum field theory constructed on a curved background spacetime experiences gravitationally induced vacuum polarization and/or particle creation. These effects induce a nonzero expectation value for the stress-energy tensor.  The renormalized expectation value of the complete quantum stress-energy tensor outside the classical event horizon has been calculated by various authors \cite{Christensen, Christensen:1976vb, Hiscock:1980ze, Hiscock:1981xb, Page, Frolov:1982fr, Elster:1983pk, Howard:1984qp, Zannias:1984tb, Howard:1985yg, Simkins, Brown:1986jy, Frolov:1987gg, Frolov:1987gw, Jensen:1988rh, Jensen:1991ef, Anderson:1993if, Anderson:1994hg, Matyjasek:1996ih, Visser,Taylor:1998dk, Matyjasek:1998mq, Groves, Carlson:2003ub, Popov:2003ne, Matyjasek:2004vd, Casals, Morgan:2007hp, Breen:2010ux, Ottewill:2010bq, Ottewill:2010hr, Breen:2011af, Breen:2011aa, Martin-Moruno:2013wfa, Belokogne:2014ysa, Levi:2015eea, Ferreira:2015cka, Levi:2016esr, Matyjasek:2016pyd, Levi:2016quh, Levi:2016exv, Levi:2016paz}, usually using a framework established by Christensen and Fulling \cite{Christensen}. In this framework, the assumptions are that the stress-energy tensor is time independent, satisfies local stress-energy conservation, and has a trace determined solely by the conformal or Weyl anomaly \cite{Capper:1974ic} for fields that are classically conformally invariant (such as the conformally coupled massless scalar field and the electromagnetic field, but not the gravitational field \cite{Critchley:1978kb}). The quantum state considered is usually either the Hartle-Hawking state \cite{Hartle} or the Unruh state \cite{Unruh0, Unruh}. (For discussions of the various quantum field theory states outside a black hole, see \cite{Birrel}.) In the Hartle-Hawking state, one has thermal equilibrium, and zero net energy flux, with the outgoing Hawking radiation balanced by incoming radiation from an external heat bath at the Hawking temperature. In \cite{Page} a fairly good closed-form approximation for the energy density and stresses of a conformal scalar field in the Hartle-Hawking state everywhere outside a static black hole can be found.

In the Unruh state, there is the absence of incoming radiation at both past null infinity and the past horizon, plus regularity of the stress-energy tensor on the future event horizon in the frame of a freely falling observer, representing a black hole formed from gravitational collapse, with nothing falling into the black hole thereafter. There have been many calculations of the quantum stress-energy tensor in the Unruh state in the Schwarzschild spacetime, both for a massless scalar field and for the electromagnetic field \cite{Elster:1983pk,Howard:1984qp,Zannias:1984tb, Howard:1985yg,Simkins,Jensen:1991ef,Anderson:1993if,Anderson:1994hg,Matyjasek:1996ih,Visser,Taylor:1998dk,Matyjasek:1998mq}.  A method for computing the stress-energy tensor for the quantized massless spin-$1/2$ field in a general static spherically symmetric spacetime was presented in \cite{Groves, Carlson:2003ub, Matyjasek:2004vd}. The quantum stress-energy tensor has also been investigated in the Kerr metric \cite{Taylor:1998dk, Casals, Belokogne:2014ysa, Ferreira:2015cka, Levi:2016exv}.

One of the important questions that one wants to answer concerns the effect of quantized matter on the geometry of black holes. Such effects in the Hartle-Hawking state have been studied \cite{York} (for similar work see \cite{HK, York3, HS}), using the approximation found in \cite{Page} for the expectation value of the renormalized thermal equilibrium stress-energy tensor of a free conformal scalar field in a Schwarzschild black hole background as the source in the semiclassical Einstein equation. The backreaction and new equilibrium metric are found perturbatively to first order in $\hbar$. The new metric is not asymptotically flat unless the system is enclosed by a reflecting wall. The nature of the modified black hole spacetime was explored in subsequent work \cite{York1, York2, HLA}.  James Bardeen \cite{Bardeen} considered radial null geodesics in a black hole geometry modified by Hawking radiation backreaction, showing that the event horizon is stable and shifted slightly in radius from the vacuum background.

In this paper, we construct the first-order backreaction on the metric in the Unruh state, using the expectation value of the quantum stress-energy tensor in the Unruh state as the source in the spherically symmetric Einstein equation. This metric represents the first-order approximation to the metric of an evaporating black hole. 
\section{metric ansatz}
To construct a metric using the expectation value of the quantum stress-energy tensor in the Unruh state as a source in the spherically symmetric Einstein equation, we first need to find an appropriate metric ansatz. To do so, we begin (but do not end) with the outgoing Vaidya metric. The outgoing Vaidya metric describes a spherically symmetric spacetime with radially outgoing null radiation. Here, we consider a black hole evaporating by Hawking emission. The outgoing Vaidya metric can be written in outgoing Eddington-Finkelstein coordinates as 
\be
ds^2=-\left(1-\frac{2\mu(u)}{r}\right)du^2-2dudr+r^2d\Omega^2 \, . \n{Vadyaur}
\ee 
Here $u$ is the retarded or outgoing null time coordinate, and $d\Omega^2=d\theta^2+\sin^2\theta d\phi^2 \,$. We use Planck units, $\hbar = c = G = 1$.

When the black hole mass $\mu$ is much larger than the reciprocal of the masses of all massive particles, the Hawking emission is almost entirely into massless particles (e.g., photons and gravitons for astrophysical mass black holes), and the Hawking emission rate is given by
 \be
\mu'\equiv\frac{d\mu}{du} \approx -\frac{\alpha}{\mu^2} \, , \n{rate}
\ee 
where $\alpha$ is a constant coefficient that has been numerically evaluated to be about $\alpha \approx 3.7474\times 10^{-5}$ \cite{Page:1976df,Page:1976ki,Page:1977um,Page:1983ug,Page:2004xp} for the emission of massless photons and gravitons. Then we have
\be
\mu(u) \approx \left(-3\alpha u\right)^{\frac{1}{3}} \, . \n{re1}
\ee 
We are setting $u = 0$ at the final evaporation of the black hole, so that $u$ is negative for the part of the spacetime being considered. Moreover, we are assuming that the black hole mass is infinite at negative infinite $u$, so going back in retarded time, the mass grows indefinitely, rather than having a black hole that forms at some particular time. However, for a black hole that forms at some initial mass $M_0$, the metric we find should be good for values of $u$ when $\mu(u) < M_0$.

For $r\gg 2\mu$, the retarded/outgoing time $u$ can be written in terms of the advanced/ingoing time $v$ and radius $r$ as approximately
\be
u \approx v-2(r-2\mu)-4\mu\ln\frac{r}{2\mu} \, . \n{u=Vr}
\ee Then $\mu(u)$ can be written in terms of $v$ and $r$ as well, since
\be
\mu^3 \approx -3\alpha v + 6 \alpha r - 12 \alpha \mu + 12 \alpha \mu \ln\frac{r}{2\mu} = -3\alpha v + 12\alpha\mu\left(\frac{1}{z} - \ln z - 1\right) \, , \n{cubiceq}
\ee
where
\be
z \equiv \frac{2\mu}{r}\, . \n{z}
\ee
In terms of the zeroth-order solution
\be
\mu_0(v) \equiv \left(-3\alpha v\right)^{\frac{1}{3}}
\n{mu_0}
\ee 
and the function
\be
\epsilon(v,z) \equiv \frac{4\alpha}{\mu_0^2}\left(\frac{1}{z} - \ln z - 1\right)
\n{epsilon}
\ee
(which vanishes on the apparent horizon at $r=2\mu$ or $z=1$), the mass $\mu$ is given approximately by the solution of the cubic equation (\ref{cubiceq}) as
\be
\mu(v,z) \approx \mu_0\left[\frac{1}{2}
+\left(\frac{1}{4}-\epsilon^3\right)^\frac{1}{2}\right]^\frac{1}{3}
+ \mu_0\left[\frac{1}{2}
-\left(\frac{1}{4}-\epsilon^3\right)^\frac{1}{2}\right]^\frac{1}{3} \, . \n{mu-cubic}
\ee

\begin{center}
\begin{figure}[t]
\centering
 \includegraphics[width=10cm]{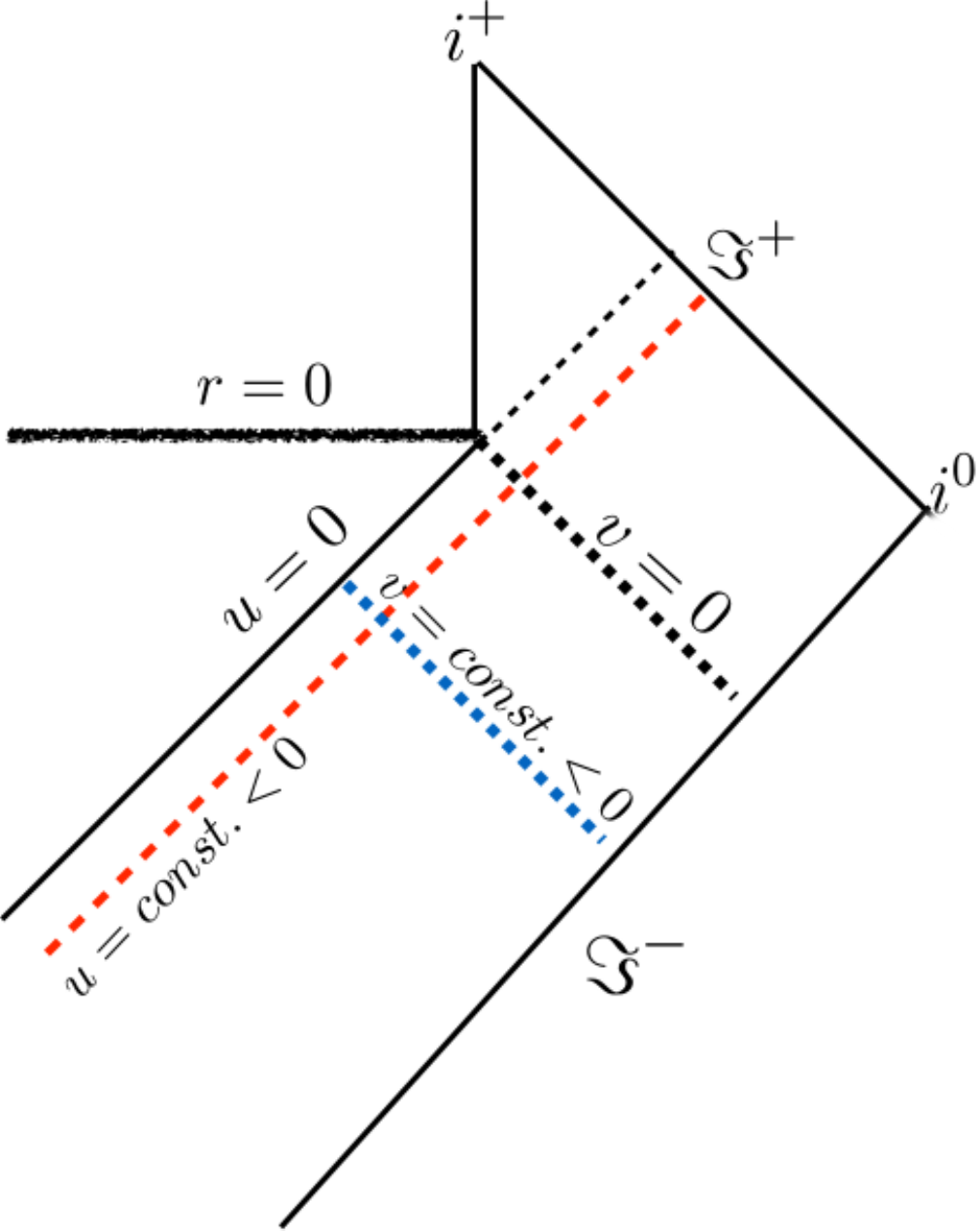}
   \caption{Carter-Penrose diagram for an evaporating black hole. The black hole mass is infinite at negative infinite $u$.  Our metric applies for $v\ll -1$ (in Planck units) and for $u<0$, though it
should also be good for $u$ somewhat positive (a bit inside the black
hole, so long as $z\equiv 2\mu/r$ is not much larger than unity).\label{CP}}
\end{figure}
\end{center}

Figure (\ref{CP}) shows a spacetime diagram for the evaporating black hole and the null coordinates $u$ and $v$.  These null coordinates are chosen so that $u = 0$ is the event horizon of the black hole, and $u = v = 0$ is the final evaporation event of the hole, when and where the horizon radius goes to $r = 0$.

We shall consider the region outside the black hole horizon, where $r > 2\mu$, so $0 < z < 1$, though our results should continue to be approximately valid somewhat inside the black hole, so long as $z \equiv 2\mu/r$ is not too much larger than unity, since the approximations used later for the stress-energy tensor were derived for $z < 1$ and are expected to become poor for $z$ significantly larger than unity, though there is no singularity or discontinuity at the horizon itself.

When $\epsilon \ll 1$, Eq.\ (\ref{mu-cubic}) implies that
\be
\mu(v,z) \approx \mu_0(1+\epsilon) = 
\left(-3\alpha v\right)^{\frac{1}{3}}+\frac{4\alpha}{\left(-3\alpha v\right)^{\frac{1}{3}}}\left(\frac{1}{z} - \ln z - 1\right) \, . \n{mu}
\ee
If $-3\alpha v\gg 1$, then Eq.\ (\ref{mu_0}) implies that $\mu_0 \gg 1$ (here 1 being the Planck mass in the Planck units we are using), the condition that the semiclassical approximation be good everywhere outside (and also somewhat inside) the black hole.  In this case, Eq.\ (\ref{epsilon}) implies that $\epsilon \ll 1$ unless $z \lesssim 4\alpha/\mu_0^2 \ll 1$, in which case $1/z \gg -\ln{z}$, so that $\epsilon \approx 4\alpha/(\mu_0^2 z) = 2\alpha r/(\mu_0^2\mu) < 2\alpha r/(\mu_0^3) = 2r/(-3v)$.  Therefore, if $r \ll -3v/2 \gg 1/\alpha$, then $\epsilon \ll 1$.  This condition is implied by $0 < -u \ll -4v \gg 4/(3\alpha)$, since then Eq.\ (\ref{u=Vr}) gives $2r \approx v-u-4\mu\ln{[r/(2\mu)]} <  -u-(-v) \ll 3(-v)$.  That is, $0 < -u \ll -4v \gg 4/(3\alpha)$ is sufficient to imply $\epsilon \ll 1$ and the validity of the approximation of Eq.\ (\ref{mu}).  (The approximation also continues to be good slightly inside the horizon, where $u > 0$.)

With $-3\alpha v\gg 1$, the second term on the right-hand side of Eq.\ (\ref{mu}) is insignificant unless $z \lesssim 4\alpha/\mu_0^2 \ll 1$, and for $4\alpha(-3\alpha v)^{-\frac{2}{3}} \ll z \ll 1$, we can take $z\approx 2\mu_0/r$, so that in any case in which $-3\alpha v\gg 1$ and $z \gg 4\alpha(-3\alpha v)^{-\frac{2}{3}}$,
\ba
\mu(v,r) &\approx& \mu_0 + 2\alpha (r-2\mu_0)/\mu_0^2
                    + (4\alpha\mu_0)\ln{[r/(2\mu_0)]} \nonumber\\
    &\equiv& \left(-3\alpha v\right)^{\frac{1}{3}}
             + 2\alpha r \left(-3\alpha v\right)^{-\frac{2}{3}} 
             -4\alpha(-3\alpha v)^{-\frac{1}{3}}
	     + 4\alpha \left(-3\alpha v\right)^{-\frac{1}{3}}\ln{[r/(-24\alpha v)^{\frac{1}{3}}]}\, . \n{muapprox}
\ea

Next, we transform the outgoing Vaidya metric to ingoing Eddington-Finkelstein coordinates with advanced/ingoing time $v$ and radius $r$. Let us consider $u$ as a function of $v$ and $r$ again as in Eq.\ (\ref{u=Vr}).  Then we have
\ba
&&du \approx dv-2dr-4\mu'\ln\frac{r}{2\mu}du-\frac{4\mu dr}{r} + 8\mu'du \nonumber\\
\Rightarrow&&du \approx e^{\tilde{\psi}(v,r)}dv-2\left(1+\frac{2\mu}{r}\right)e^{\tilde{\psi}(v,r)}dr \, , \n{du}
\ea 
where
\ba
&&e^{-\tilde{\psi}(v,r)}\equiv1-8\mu'+4\mu' \ln\frac{r}{2\mu} \, .
\ea 
For $0 < -u \ll -4v \gg 4/(3\alpha)$ so that $\epsilon \ll 1$ and hence $\mu-\mu_0 \ll \mu_0$,
\ba
&&\tilde{\psi}(v,r) \approx -4\mu'\left(\ln\frac{r}{2\mu}-2\right) \approx -\frac{4\alpha}{\mu^2}\left(\ln z+2\right)\, .  \n{tildepsi}
\ea 
Plugging back Eq.\ (\ref{du}) into the metric (\ref{Vadyaur}), we get
\ba
ds^2&\approx&-\left(1-\frac{2\mu}{r}\right)e^{2\tilde{\psi}(v,r)}dv^2 +2e^{\tilde{\psi}(v,r)}\left[2\left(1+\frac{2\mu}{r}\right)\left(1-\frac{2\mu}{r}\right)e^{\tilde{\psi}(v,r)}-1\right]dvdr \nonumber\\
&+&4\left(1+\frac{2\mu}{r}\right)e^{\tilde{\psi}(v,r)} \left[1-\left(1+\frac{2\mu}{r}\right) \left(1-\frac{2\mu}{r}\right)e^{\tilde{\psi}(v,r)}\right]dr^2
+r^2d\Omega^2 \, . \n{Vadya1}
\ea 

For $1 \ll 2\mu_0 \ll r \ll -3v/2$, we thus have 
\be
ds^2\approx-\left(1-\frac{2\mu}{r}\right)e^{2\tilde{\psi}(v,r)}dv^2+2e^{\tilde{\psi}(v,r)} dr dv+ r^2d\Omega^2. \n{metricS}
\ee

When $2\mu_0 \ll r$ is not true, one is close enough to the black hole that the small deviations of the metric from the Schwarzschild metric are not well approximated by the Vaidya metric, since the stress-energy tensor is not well approximated by purely outgoing null radiation as it would be in the Vaidya metric.  However, for $1 \ll 2\mu_0$, the metric is close to a slowly varying Schwarzschild metric, so it is an excellent approximation to solve the Einstein equation as a deviation from Schwarzschild to linear order in the stress-energy tensor of the Hawking radiation, which itself can be well approximated by what it would be for a fixed Schwarzschild metric background.  Furthermore, when one goes out from the black hole along the past-directed null direction with $v =$ const.\ $\ll -1$, the energy of the Hawking radiation that one crosses increases the mass so that it is no longer well approximated by $\mu_0(v) \equiv (-3\alpha v)^{\frac{1}{3}}$, as $z \equiv 2\mu/r$ becomes sufficiently small that $\epsilon(v,z) \equiv (4\alpha/\mu_0^2)(z^{-1}-\ln{z}-1)$ does not remain small.  Then Eq.\ (\ref{tildepsi}) is no longer a good approximation for what $\tilde{\psi}(v,r)$ should be.

Therefore, we modify the metric (\ref{metricS}) with corrections going as $1/\mu^2$ with coefficients going mainly as functions of $z$ in order to match the stress-energy tensor of the Hawking radiation under the approximation that it is produced by a slowly varying sequence of Schwarzschild metrics.  In particular, we write the general spherically symmetric metric in ingoing Eddington-Finkelstein coordinates as
\be
ds^2= -\, e^{2\psi(v,z)}\left(1-\frac{2m(v,z)}{r}\right)dv^2+2e^{\psi(v,z)} dvdr+r^2 d\Omega^2\, , \n{an1}
\ee
and, motivated by the approximate forms above for the variables $\mu$ and $\tilde{\psi}$ in the metric (\ref{metricS}), we make the following ansatz for the metric coefficients of the evaporating black hole: 
\ba
&&\psi\approx \frac{1}{\mu_0^2}\left[g(z) - 4\alpha \ln{\tilde{z}}\right],\n{psirelation}\\
&&m \approx \mu\left[1 + \frac{h(z)}{\mu_0^2}\right] \n{m}, 
\ea
where $\mu(v,z)$ is given by Eq.\ (\ref{mu-cubic}), or approximately by Eq.\ (\ref{mu}) for $2r \ll -3v$, $z \equiv 2\mu/r$, and $\tilde{z}$ is a function of $\mu/\mu_0$ that is approximately $z$ for $2r \ll -3v$ but whose form for larger $r$ will be evaluated later.

Once we find the functions $g(z)$, $h(z)$, and $\tilde{z}(\mu/\mu_0)$ to solve the Einstein equation for the small deviation of the metric from a sequence of Schwarzschild metrics with a slowly decreasing mass, the resulting metric (\ref{an1}) should be a good approximation for the exact nonrotating black hole metric for all $v \ll -1$ (that is, for all advanced times before the black hole shrinks down to near the Planck mass) and for $z \equiv 2\mu/r \lesssim 1$ (that is, for all radii outside and even slightly inside the black hole), since the mass will be changing at a timescale much longer than the timescale of the mass itself.  For $r \gg 1$ (radii large in Planck units), the metric should also be good for $v \gtrsim -1$ [at advanced times after the black hole shrinks down to near the Planck mass, where the metric has the approximate outgoing Vaidya form of Eq.\ (\ref{Vadyaur})].  That is, the metric (\ref{an1}) with $\psi$ given by Eq.\ (\ref{psirelation}) and $m$ given by Eq.\ (\ref{m}), with $g(z)$, $h(z)$, and $\tilde{z}(\mu/\mu_0)$ to be found below, should be a good approximation for the metric everywhere outside, and even slightly inside, a black hole evaporating from infinite mass, except in the spacetime region near the final evaporation event of the black hole and to the causal future of this region (where one would expect quantum gravity effects to be important).

In summary, we started with the outgoing Vaidya metric as a first approximation for a spherically symmetric black hole metric evaporating by the emission of massless Hawking radiation, and then we switched to ingoing Eddington-Finkelstein coordinates and introduced the functions $g(z)$ and $h(z)$ to allow us to get a better approximation for the metric of an evaporating black hole with the stress-energy tensor of massless Hawking radiation, valid everywhere outside the black hole (and also slightly inside) that is not to the causal future of the Planckian region where the black hole has shrunk to near the Planck mass.

For $z=1$, we shall make the gauge choice of setting $g(1) = h(1) = \ln{\tilde{z}(1)} = 0$, so that there $2m \approx 2\mu = r$ and $\psi \approx 0$.  $z=1$ is then the approximate location of the apparent horizon. For $z\rightarrow 0$ (radial infinity backwards along the ingoing radial null curve of fixed $v$), we have $g(z)$, $h(z)$, and $\tilde{z}$ approaching the constants $g_0$, $h_0$, and $\tilde{z}_0$ respectively, so in this limit of infinitely large $r/\mu$, $m/\mu \rightarrow 1 + h_0/\mu_0^2$ and $\psi\rightarrow (g_0 - 4\alpha\ln{\tilde{z}_0})/\mu_0^2$.  It is only an approximation that $\psi$ and $m$ are functions just of $\mu$ and $z$ of this form, but for a large and hence very slowly evaporating black hole, it seems to be a very good approximation.

For $r \ll -v = \mu_0^3/(3\alpha)$, which implies that $\mu \approx \mu_0 \equiv (-3\alpha v)^{\frac{1}{3}}$, Eqs.\ (\ref{rate}) and (\ref{m}) above show that at fixed $z$ the mass of the black hole changes as 
\be
\frac{dm}{dv}\approx -\, \frac{\alpha}{\mu^2}.
\ee

From Eqs.\ (\ref{z})-(\ref{mu-cubic}) and (\ref{an1})-(\ref{m}), we get that the area of the apparent horizon (at $z=1$) of the evaporating black hole is
\be
A(v) = 4\pi\left[2m(v,r=2m)\right]^2 \approx 4\pi\left[2\mu(v,z=1)\right]^2 =16\pi{\mu_{0}}^2
  = 16\pi(-3\alpha v)^\frac{2}{3}  \, . \n{A}
\ee

Note that Bardeen \cite{Bardeen} considers a quasistationary approximation of the black hole, which is justified as long as the black hole mass is much larger than the Planck mass $m_p\equiv (\hbar c/G)^{1/2}$, which we are setting equal to unity by using Planck units.  In this case Bardeen has the black hole mass at $r\sim 2\mu_0$, $M\equiv \mu(v,{r=2\mu_0})$, being proportional to $(-v)^{1/3}$. 

The metric (\ref{an1}) can be written in the form
\be
ds^2= -\, e^{2\Psi}\left(1-\frac{2\tilde{M}}{r}\right)dv^2+2e^{\Psi} dvdr+r^2 d\Omega^2\, ,
\ee where $\Psi\equiv\Psi(v,r)=\psi(v,z=2\mu/r)$ and $\tilde{M}\equiv\tilde{M}(v,r)=m(v,z=2\mu/r)$. Components of the Einstein tensor for this metric have the following form:
\ba
{G^{v}}_{v}&=&-\, \frac{2}{r^2}\tilde{M}_{,r}\, , \\
{G^{r}}_{v}&=&\frac{2}{r^2}\tilde{M}_{,v}\, ,\\
{G^{v}}_{r}&=&\frac{2}{r} e^{-\Psi}\Psi_{,r}\, , \\
{G^{r}}_{r}&=&\left(1-\frac{2\tilde{M}}{r}\right)\frac{2\Psi_{,r}}{r}-\frac{2\tilde{M}_{,r}}{r^2}\, , \\
{G^{\phi}}_{\phi}={G^{\theta}}_{\theta}&=&\left(1-\frac{2\tilde{M}}{r}\right)\left(\Psi_{,r}^{2}+\Psi_{,rr}\right)+\frac{1}{r}\left(1+\frac{\tilde{M}}{r}-3\tilde{M}_{,r}\right)\Psi_{,r}-\frac{1}{r}\tilde{M}_{,rr}+e^{-\Psi}\Psi_{,rv}\, .
\ea

Taking $v$ and $z$ to be the independent coordinates (at fixed angles $\theta$ and $\phi$), we can rewrite the metric (\ref{an1}) for $2r \ll -3v$ [the region where $g(z)$, $h(z)$, and $\tilde{z}$ have significant variation; outside this region we can set $g \approx g_0$, $h \approx h_0$, and $\tilde{z} \approx \tilde{z}_0$ in the metric (\ref{an1})] as
\ba
ds^2&\approx&-\biggl[e^{2\psi(v,z)}\left(1-z\left(1+\frac{h(z)}{{\mu_{0}}^2}\right)\right)-2e^{\psi(v,z)}r_{,v}\biggl]dv^2+2e^{\psi(v,z)}r_{,z}dzdv +\frac{4\mu^2}{z^2}d\Omega^2\nonumber\\
 &=&-A dv^2+2B dvdz+r^2d\Omega^2 \, , \n{metric2}
\ea where
\ba
 &&r_{,v}\approx -\, \frac{2\alpha}{z\left(-3\alpha v\right)^{\frac{2}{3}}}+\frac{8\alpha^2}{z^2}\, \frac{1-z\ln z-z}{\left(-3\alpha v\right)^{\frac{4}{3}}} \, ,\n{rveq} \\
 && r_{,z}\approx -\, \frac{2}{z^2}\left(-3\alpha v\right)^{\frac{1}{3}}-\frac{8\alpha}{z^3} \, \frac{2-z\ln z}{\left(-3\alpha v\right)^{\frac{1}{3}}}\,\n{rzeq} .
\ea

Although this form of the metric does not apply everywhere outside a large black hole as the metric (\ref{an1}) does, it applies where the stress-energy tensor of the Hawking radiation gives significant contributions to the functions $g(z)$, $h(z)$, and $\tilde{z}$ in Eqs.\ (\ref{psirelation}) and (\ref{m}).  Therefore, this form of the metric will be used to get approximate solutions of the Einstein equation for $g(z)$, $h(z)$, and $\tilde{z}$ with the stress-energy tensor of the Hawking radiation of massless fields in the Unruh quantum state.  Then these will be inserted back into Eqs.\ (\ref{psirelation}) and (\ref{m}) to give the functions in the metric (\ref{an1}), which will be a good approximation for the metric everywhere outside a large black hole evaporating by Hawking radiation of massless fields.

The stress-energy tensor for the spherically symmetric Unruh quantum state $\left|\psi\right>$ on the spherically symmetric curved background of the Schwarzschild spacetime with $2\mu \equiv rz$ constant,
\ba
&&ds^2= -\, (1-z)dt^2+\frac{dr^2}{1-z}+r^2d\Omega^2 \, ,
\ea in the standard static orthonormal frame
\ba
&&\omega^{\hat{0}}=\sqrt{1-z} \, dt,\\
&&\omega^{\hat{1}}=\frac{1}{\sqrt{1-z}} \, dr,\\
&&\omega^{\hat{2}}=r \, d\theta,~~\omega^{\hat{3}}=r\sin\theta \, d\phi,
\ea can be written in the following form:
\ba
\left\langle\psi\right| {T}^{\hat{\mu}\hat{\nu}}\left|\psi\right>=
 \begin{pmatrix}
 \tilde{\rho} & \tilde{f} & 0 & 0 \\
  \tilde{f}  & \tilde{P}  & 0 & 0 \\
  0 & 0 & \tilde{p} & 0\\
  0 & 0 & 0 & \tilde{p}
 \end{pmatrix}.\n{EMT}
 \ea 
 
Note that we are using a capital $\tilde{P}$ for the radial pressure and a lowercase $\tilde{p}$ for the transverse pressure. Dimensional analysis shows that for massless fields at fixed $z$, the dependence of the orthonormal components of the stress-energy tensor on the mass $\mu$ of the Schwarzschild metric at fixed $z$ goes as $\mu^{-4}$, so for the slowly evolving metric ({$\ref{metric2}$}), we shall assume that the stress-energy tensor ({$\ref{EMT}$}) has approximately the following form with functions $\rho(z)$, $f(z)$, $P(z)$, and $p(z)$ that are dimensionless even without setting $G=1$ (and hence independent of the scale set by $\mu_0$):
\ba
\left\langle\psi\right| T^{\hat{\mu}\hat{\nu}}\left|\psi\right>=\frac{1}{\mu_0^4}
 \begin{pmatrix}
 \rho(z) & f(z) & 0 & 0 \\
  f(z)  & P(z)  & 0 & 0 \\
  0 & 0 & p(z) & 0\\
  0 & 0 & 0 & p(z)\n{EMT2}
 \end{pmatrix}. 
 \ea

According to Christensen and Fulling \cite{Christensen}, the stress-energy tensor in the case of the Schwarzschild spacetime can be decomposed into four separately conserved quantities. We follow the analysis of Matt Visser \cite{Visser} and use his form of the stress-energy tensor, which has a slightly different basis for its decomposition from that of Christensen and Fulling and is given by 
 \be
 \left\langle\psi\right| T^{\hat{\mu}\hat{\nu}}\left|\psi\right>= [T_{\text {trace}}]^{\hat{\mu}\hat{\nu}}+[T_{\text{pressure}}]^{\hat{\mu}\hat{\nu}}+[T_{+}]^{\hat{\mu}\hat{\nu}}+[T_{-}]^{\hat{\mu}\hat{\nu}} \, , 
 \ee
where, with $T(z)$, $H(z)$, and $G(z)$ being a further set of functions of $z \equiv 2\mu/r$ that are dimensionless without setting the Newtonian gravitational constant to be unity, and with $f_+$ and $f_-$ being dimensionless constants,
\ba
 \mu_0^4[T_{\text{trace}}]^{\hat{\mu}\hat{\nu}}\equiv  \begin{pmatrix}
 -T(z)+\frac{z^2}{1-z}H(z) & 0 & 0 & 0 \\
  0 & \frac{z^2}{1-z}H(z)  & 0 & 0 \\
  0 & 0 & 0 & 0\\
  0 & 0 & 0 & 0
 \end{pmatrix},\n{THdecomposition}
\ea
 \ba
H(z)&&\equiv \frac{1}{2}\int_{z}^{1}\frac{T(\bar{z})}{\bar{z}^2}d\bar{z} \,\n{HT1} ,
\ea 
\ba
 \mu_0^4[T_{\text{pressure}}]^{\hat{\mu}\hat{\nu}}\equiv  \begin{pmatrix}
 2p(z)+\frac{z^2}{1-z}G(z) & 0 & 0 & 0 \\
  0 & \frac{z^2}{1-z}G(z)  & 0 & 0 \\
  0 & 0 & p(z) & 0\\
  0 & 0 & 0 & p(z)
 \end{pmatrix},\n{TGdecomposition}
\ea
\ba
G(z)&&\equiv\int_{z}^{1}\left[\frac{2}{\bar{z}^3}-\frac{3}{\bar{z}^2}\right]p(\bar{z})d\bar{z} \, \n{Gp1},
\ea 
\ba
 \mu_0^4[T_{+}]^{\hat{\mu}\hat{\nu}}\equiv  f_{+}\frac{z^2}{1-z}\begin{pmatrix}
 1 & 1 & 0 & 0 \\
  1 & 1  & 0 & 0 \\
 0  & 0  & 0 & 0\\
  0 & 0 & 0 & 0
 \end{pmatrix} \, ,\n{Tplus}
 \ea 
\ba
 \mu_0^4[T_{-}]^{\hat{\mu}\hat{\nu}}\equiv  f_{-}\frac{z^2}{1-z}\begin{pmatrix}
 1 & -1 & 0 & 0 \\
  -1 & 1  & 0 & 0 \\
 0  & 0  & 0 & 0\\
  0 & 0 & 0 & 0
 \end{pmatrix} \, ,\n{Tminus}
 \ea 

The decompositions (\ref{THdecomposition}) and (\ref{TGdecomposition}) make sense if the integrals $G(z)$ and $H(z)$ converge. Imposing mild integrability constraints on $T(z)$ and $p(z)$ at the apparent horizon that is very near $z=1$, which are satisfied for the Unruh state where $T(1)$ and $p(1)$ are actually finite, we have
 \ba
&&H(z)=\frac{1}{2} T(1) (1-z)+O[(1-z)^2],\n{Hequation}\\
&&G(z)= -\, p(1) (1-z)+O[(1-z)^2].\n{Gequation}
\ea
This is enough to imply that the two tensors (\ref{THdecomposition}) and (\ref{TGdecomposition}) are individually regular at both the past and future horizon. 
 In Kruskal null coordinates, $[T_{+}]$ is singular on the future horizon $H^+$ and regular on the past horizon $H^-$. On the other hand, $[T_{-}]$ is singular on the past horizon $H^-$ and regular on the future horizon $H^+$. The two tensors (\ref{Tplus}) and (\ref{Tminus}) correspond to outgoing and ingoing null fluxes, respectively. The constants $f_{+}$ and $f_{-}$ determine the overall flux. The Unruh state must be regular on the future horizon, so we need $f_{+}=0$. However, such a condition na\"{\i}vely seems to exclude any outgoing radiation. Nevertheless, we can get outgoing radiation by making $f_{-}$ negative. It is convenient to define $\beta$ to be what $p(0)=\mu^4\tilde{p}(r=\infty)$ would be for massless scalar radiation in the Hartle-Hawking state at large radii (ignoring the backreaction to the Schwarzschild metric),
 \be
 \beta\equiv \frac{1}{2^{13} 3^2 5 \pi^2}\equiv \frac{1}{368\,640\pi^2}\, ,
\ee and set $f_{-}=-\beta f_{0}$, where $f_{0}$ is a positive quantity.
In what follows, we also define 
 \ba
 f(z)\equiv \beta f_0\frac{z^2}{1-z}. \n{fzrelation}
 \ea 
Thus,
\ba
 \mu_0^4[T_{+}]^{\hat{\mu}\hat{\nu}}+\mu_0^4[T_{-}]^{\hat{\mu}\hat{\nu}}\equiv \begin{pmatrix}
 -f(z) & f(z) & 0 & 0 \\
  f(z) & -f(z)  & 0 & 0 \\
 0  & 0  & 0 & 0\\
  0 & 0 & 0 & 0
 \end{pmatrix} \, .
 \ea 

Therefore, we have
\ba
\mu_0^4T^{\hat{0}\hat{0}}&=&\rho(z)=2p(z)+\frac{z^2}{1-z}\left[H(z)+G(z)\right]-T(z)-f(z)\, ,\n{EMW1} \\
\mu_0^4T^{\hat{1}\hat{0}}&=&\mu_0^4T^{\hat{0}\hat{1}}=f(z) \, , \\
\mu_0^4T^{\hat{1}\hat{1}}&=&P(z)= \frac{z^2}{1-z}\left[H(z)+G(z)\right] - f(z) \, , \\
\mu_0^4T^{\hat{2}\hat{2}}&=&\mu_0^4T^{\hat{3}\hat{3}}=p(z) \n{EMW4} \, .
\ea 

Since the stress-energy tensor is given in the orthonormal frame, we rewrite the Einstein equation in the orthonormal frame, i.e., 
\be
G_{\hat{\mu}\hat{\nu}} = 8\pi T_{\hat{\mu}\hat{\nu}}. \n{EinsteinEq}
\ee
To find the approximate time-dependent metric for an evaporating black hole as a first-order perturbation of the Schwarzschild metric, using the linearized backreaction from the stress-energy tensor (\ref{EMW1})-(\ref{EMW4}) of the Hawking radiation in the Unruh quantum state in the Schwarzschild spacetime, we solve Eq.\ (\ref{EinsteinEq}) up to relative corrections of the order of $1/\mu^2$ in the Planck units that we are using.  To bring $G_{\mu\nu}$ to the orthonormal frame, note that we have
\ba
ds^2 = -\, (\omega^{\hat{0}})^2+(\omega^{\hat{1}})^2+(\omega^{\hat{2}})^2+(\omega^{\hat{3}})^2=-\, A dv^2+2B dvdz+r^2d\Omega^2,
\ea
where 
\ba
&&\omega^{\hat{0}}=\sqrt{A}dv-\frac{B}{\sqrt{A}}dz,\\
&&\omega^{\hat{1}}=\frac{B}{\sqrt{A}}dz,\\
&&\omega^{\hat{2}}=rd\theta,~~\omega^{\hat{3}}=r\sin\theta d\phi. 
\ea
We have that the Einstein tensor is
\ba
G&=&G_{\hat{0}\hat{0}}\omega^{\hat{0}} \omega^{\hat{0}}+G_{\hat{0}\hat{1}}\omega^{\hat{0}} \omega^{\hat{1}}+G_{\hat{1}\hat{0}}\omega^{\hat{1}} \omega^{\hat{0}}+
G_{\hat{1}\hat{1}}\omega^{\hat{1}} \omega^{\hat{1}}+G_{\hat{2}\hat{2}}\omega^{\hat{2}}\omega^{\hat{2}}+
G_{\hat{3}\hat{3}}\omega^{\hat{3}}\omega^{\hat{3}}\nonumber\\
&=&G_{vv}dv^2+2G_{vz} dvdz+G_{zz}dz^2+G_{\theta\theta}d\theta^2+G_{\phi\phi}d\phi^2. 
\ea
Therefore, we obtain
\ba
&&G_{\hat{0}\hat{0}}= \frac{G_{vv}}{A},\n{OrthG1}\\
&&G_{\hat{0}\hat{1}}=G_{\hat{1}\hat{0}}=\frac{G_{vz}}{B}+\frac{G_{vv}}{A},\\
&&G_{\hat{1}\hat{1}}=\frac{AG_{zz}}{B^2}+2\frac{G_{vz}}{B}+\frac{G_{vv}}{A},\\
&&G_{\hat{2}\hat{2}} =\frac{G_{\theta\theta}}{r^2}, \\
&&G_{\hat{3}\hat{3}} =\frac{G_{\phi\phi}}{r^2\, \sin^2\theta}.
\n{OrthG}
\ea 
Deriving the Einstein tensor components $G_{vv}$, $G_{vz}$, $G_{zz}$, and $G_{\theta\theta}$ for the metric ($\ref{metric2}$) and using Eqs.\ (\ref{OrthG1})-(\ref{OrthG}), we obtain the Einstein tensor components in the orthonormal frame.

\section{Solution for the metric coefficients}

We now solve the Einstein equation (\ref{EinsteinEq}) to first order in the perturbation of the metric from the Schwarzschild metric, using the stress-energy tensor whose components are proportional to $1/\mu^4$.  From the zero-one component of the Einstein equation, we get
\be
f(z)=\frac{\alpha z^2}{16 \, \pi \, (1-z)} \, \n{funf}.
\ee
Comparing this to Eq.\ (\ref{fzrelation}), we have $f_0=\alpha/(16\pi\beta)$. Therefore, the Hawking radiation luminosity of the black hole is not only $L = - dm/dv \approx \alpha/\mu^2$ but also $16\pi \beta f_0/\mu^2$. From the zero-zero component, we find
\be
\rho(z)=\frac{z^2}{32 \, \pi \, (1-z)}\left[{2\alpha\left(1-2z^2\right)-z^2(1-z)h_{,z}}\right]=f(z)\left(1-2z^2\right)-\frac{z^4 \, h_{,z}}{32\pi}\, \n{rhofromhg}.
\ee 
From the one-one component of the Einstein equation (\ref{EinsteinEq}), we get
\ba
P(z)&=&\frac{z^2}{32 \, \pi \, (1-z)}\left[{2\alpha\left(1-8z+6z^2\right)-2z(1-z)^2g_{,z}+z^2(1-z)h_{,z}}\right] \nonumber\\
&=&f(z)\left(1-8z+6z^2\right)-\frac{1}{16\pi} z^3 (1-z) g_{,z}+\frac{1}{32\pi} z^4 h_{,z} \, \n{taufromhg}.
\ea 
From the two-two component, we get
\ba
p(z)&=&\frac{z^3}{64 \, \pi  }\left[16\alpha+(2-5z)\, g_{,z}-2zh_{,z}+2\, z\, (1-z)g_{,zz}-z^2 \, h_{,zz}\right] \nonumber\\ 
&=& 4 f(z) \, z \, \left(1-z\right)+\frac{z^3}{64 \, \pi  }\left[(2-5z)\, g_{,z}-2zh_{,z}+2\, z\, (1-z)g_{,zz}-z^2 \, h_{,zz}\right]\, .
\ea

Now let us suppose that in the stress-energy tensor components (\ref{EMW1})-(\ref{EMW4}), the functions $p(z)$ and $T(z)$ are explicitly given. We then solve for the metric functions $h(z)$ and $g(z)$ in terms of $p(z)$ and $T(z)$. Moreover, we are making the gauge choice of setting $g(1)=h(1)=0$. From Eq.\ (\ref{funf}), we have $f(z)$. 

From $G_{\hat{0}\hat{0}} = 8\pi \, T_{\hat{0}\hat{0}}$ and $G_{\hat{1}\hat{1}} = 8\pi \, T_{\hat{1}\hat{1}}$, we get
\ba
g(z)&=&-\int^1_z \left[- \,\frac{2\alpha (1-4\bar{z}+2\bar{z}^2)}{\bar{z}(1-\bar{z})^2}+16\pi\frac{P(\bar{z})+\rho(\bar{z})}{\bar{z}^3(1-\bar{z})}\right]d\bar{z} \,  \nonumber\\
&=&-\int^1_z \left[32\pi\frac{ H(\bar{z})+G(\bar{z})}{\bar{z}(1-\bar{z})^2} -16\pi \frac{T(\bar{z})-2p(\bar{z})}{\bar{z}^3(1-\bar{z})} -\frac{4\alpha}{\bar{z}}\right]d\bar{z}\, , \\ \n{gfunction} 
h(z)&=&\int^1_z \left[ \frac{2\alpha(1-2\bar{z}^2)}{\bar{z}^2 \, (1-\bar{z})}-\frac{32\pi \, \rho(\bar{z})}{\bar{z}^4}\right] d\bar{z}\,  \, \nonumber\\
&=&\int^1_z \left[-\,32\pi\frac{ H(\bar{z})+G(\bar{z})}{\bar{z}^2(1-\bar{z})} +32\pi \frac{T(\bar{z})-2p(\bar{z})}{\bar{z}^4}+4\alpha\frac{(1+\bar{z})}{\bar{z}^2}\right]d\bar{z}\, . \n{hfunction}
\ea

Note that the functions $H(z)$ and $G(z)$ are given in terms of $T(z)$ and $p(z)$ by Eqs.\ (\ref{HT1}) and (\ref{Gp1}). For any conformally invariant quantum field, the trace of the stress tensor is known exactly and is given by the conformal anomaly. In the Schwarzschild spacetime, the dimensionless trace is $T(z) = \mu^4 T_\alpha^\alpha$, where
\be
T(z) = \gamma z^6 = \beta \xi z^6 \,, \n{Tracerelation}
\ee 
with 
$\beta\equiv 1/(2^{13} 3^2 5 \pi^2)$. 
Here $\xi$ is a dimensionless coefficient of the trace. For spins 0, $1/2$, 1, $3/2$, and 2 (with particles identical to antiparticles, so that for each momentum, there is a single one-particle state for spin 0 and two for higher spins, one for each of the two helicities), $\xi$ is, respectively, $96$, $168$, $-1248$, $-5592$, and $20352$ \cite{Birrel}, but this does not give the full trace for gravitons, which are not conformally invariant \cite{Critchley:1978kb}. 
Equation ($\ref{HT1}$) then gives
\ba
H(z)=\frac{\gamma}{10}(1-z^5)=\frac{\beta\xi}{10}(1-z^5)\,. 
\ea
Therefore, we can derive $g(z)$ and $h(z)$ using these special forms for $T(z)$ and $H(z)$:
\ba
g(z)&=&32\pi\int^1_z \left(\frac{G(\bar{z})}{\bar{z}(1-\bar{z})^2} + \frac{p(\bar{z})}{\bar{z}^3(1-\bar{z})} \right)d\bar{z}+\frac{8}{15}\pi \gamma \left(1-z\right)\left(29+17z+8z^2\right)+(4\alpha-\frac{16}{5}\pi\gamma) \ln{z}\,,\\
h(z)&=&32\pi \int^1_z \frac{1}{\bar{z}^4}\left(2p(\bar{z})+\frac{\bar{z}^2}{1-\bar{z}}G(\bar{z})\right)d\bar{z}+\frac{8}{5}\pi \gamma \frac{(1-z)}{z}(2-3z-5z^2-6z^3)+\left(4\alpha-\frac{16}{5}\pi\gamma\right)\ln z-\frac{4\alpha}{z}(1-z)\,.\nonumber\\
\ea
\begin{center}
\begin{figure}[t]
\centering
  \includegraphics[width=7cm]{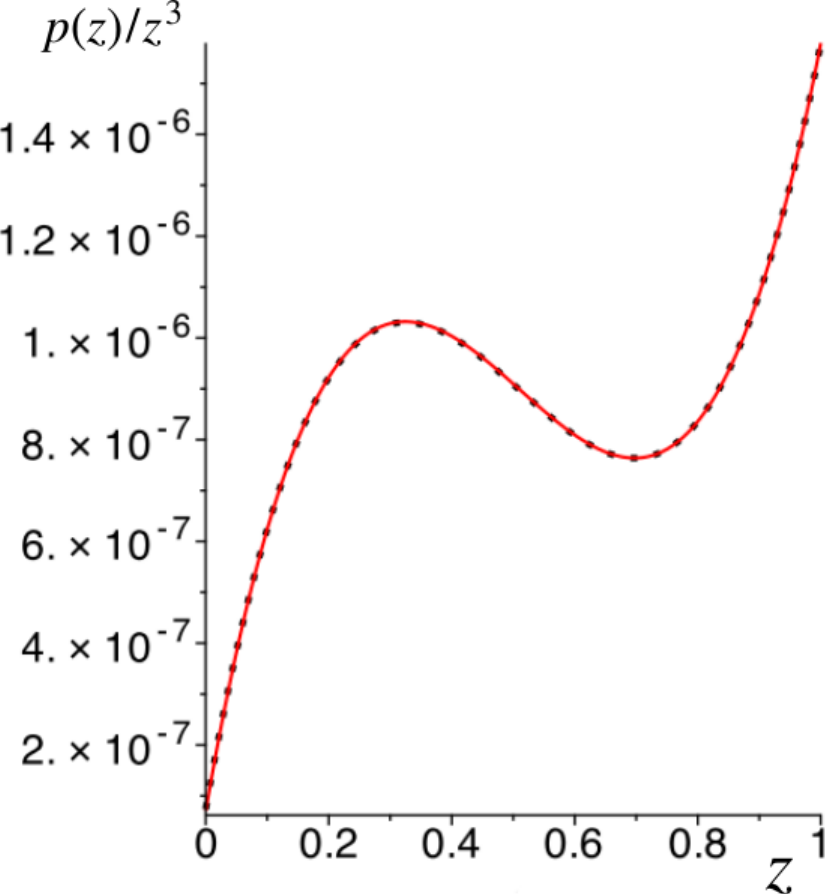}~~~~~~~
   \includegraphics[width=7cm]{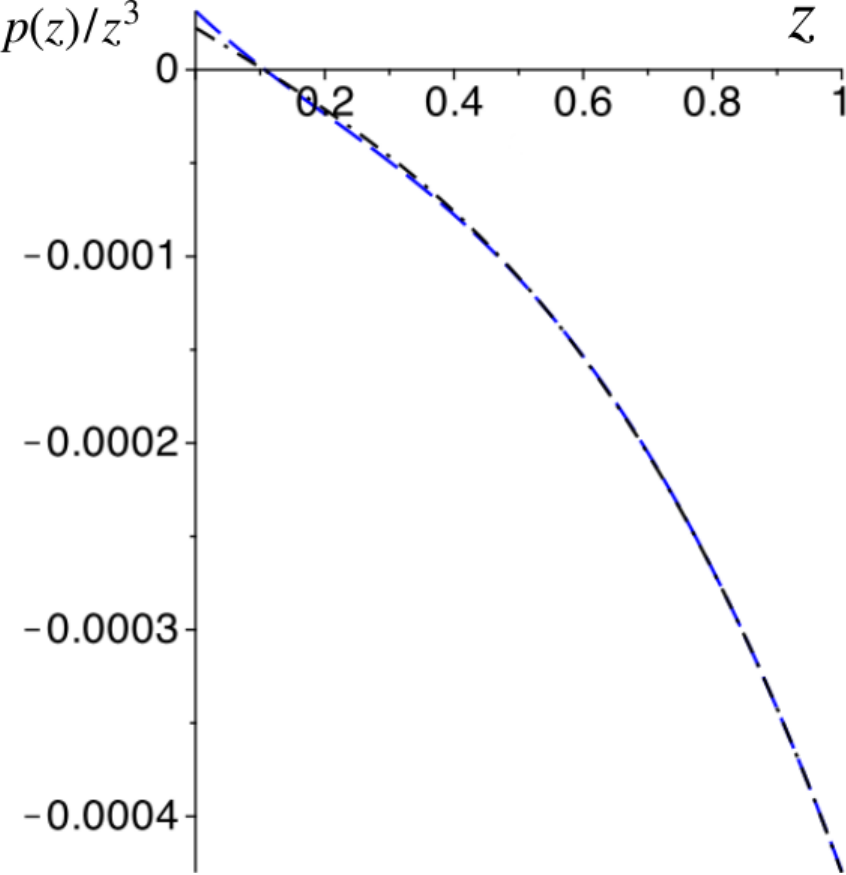}
   \caption{(a) Behaviour of the function $p(z)/z^3$ for spin 0, black dotted line for the case of nonzero $k_3$, $k_4$, $k_5$, and $k_6$ given by \cite{Bardeen:2017ypp} and red solid line for the case of nonzero $k_3$, $k_4$, $k_5$, and $k_6$ given by \cite{Bardeen:2018omt}. (b) Behaviour of the function $p(z)/z^3$ for spin 1, blue dashed line for the case of nonzero $k_3$, $k_4$, $k_5$, $k_6$, and $k_7$ given by \cite{Bardeen:2017ypp} and black dash-dotted line for the case of nonzero $k_3$, $k_4$, $k_5$, and $k_6$ given by \cite{Bardeen:2018omt}. \label{P}}
\end{figure}
\end{center}

\vspace{-0.6cm} 

A five-term polynomial is believed to be a good approximation for the function $p(z)$.  There is evidence \cite{Bardeen:2017ypp, Bardeen:2018omt} that $p(z)$ may start off at order $z^3$ (but see below), and the anomalous trace introduces a term of $z^6$. Also, for spin-1 particles, there exists evidence \cite{Bardeen:2017ypp} for a $z^7$ term. Therefore, we consider $p(z)$ to be a polynomial of the form 
\be
p(z)=\beta ({k_3z^3}+k_4z^4+k_5 z^5+k_6 z^6+k_7z^7)\,. \n{pform}
\ee 

Figure (\ref{P}) gives graphs of $p(z)/z^3$ from numerical values of the $k_i$ coefficients given later, below Eq.\ (\ref{Chg}).  Using Eq.\ ($\ref{pform}$) for $p(z)$ in Eq.\ (\ref{Gp1}) gives
\ba
G(z) &=& \beta(1-z)\left[\frac{k_3}{2}(1-3z)-k_4 z^2 -\frac{k_5}{12}(1+z+z^2+9z^3)\right.\nonumber\\
&&\left. -\, \frac{k_6}{10}(1+z+z^2+z^3+6z^4) -\, \frac{k_7}{10}(1+z+z^2+z^3+z^4+5z^5)\right]\,.
\ea

Finally, we can write $f(z)$,  $\rho(z)$ and $P(z)$ in the following form: 
\ba
f(z)&=&\beta f_0\frac{z^2}{1-z} \, ,\n{fequation} \\
\rho(z)&=&-\, f(z)+\beta z^2\left[\frac{\xi}{10} (1+z+z^2+z^3-9z^4)+\frac{k_3}{2}(1+z)+k_4z^2
-\frac{k_5}{12}(1+z+z^2-15z^3)\right.\nonumber\\
&&\ \ \ \ \ \ \ \ \ \ \ \ \ \ \ \ \ \ \ \ \ \ \ \ \ \ \ \ \ \ -\left.\frac{k_6}{10}(1+z+z^2+z^3-14z^4)-\frac{k_7}{10}(1+z+z^2+z^3+z^4-15z^5)\right],
\n{rhoequation} \\
P(z)&=& -\, f(z)+\beta z^2 \left[\frac{\xi}{10}(1+z+z^2+z^3+z^4)+\frac{k_3}{2}(1-3z)
-k_4z^2-\, \frac{k_5}{12}(1+z+z^2+9z^3)\right.\nonumber\\
&&\left.\ \ \ \ \ \ \ \ \ \ \ \ \ \ \ \ \ \ \ \ \ \ \ \ \ \ \ \ \ \ \ -\frac{k_6}{10}(1+z+z^2+z^3+6z^4)-\frac{k_7}{10}(1+z+z^2+z^3+z^4+5z^5)\right]. \n{tauequation}
\ea

We want at asymptotic infinity the stress-energy tensor to be that of an outgoing flux of positive radiation, requiring $\rho(z)\rightarrow f(z)$ asymptotically as $z\rightarrow 0$ \cite{Christensen}. Picking up the dominant terms $[O(z^2)]$ in (\ref{fequation}) and (\ref{rhoequation}), we see that    
\ba
&&f_0=\frac{\xi}{20}+{\frac{k_3}{4}}-\frac{k_5}{24}-\frac{k_6}{20}-\frac{k_7}{20}\n{7}\,.
\ea 
In the case that $k_3=0$, this constraint is the same as Eq.\ (29) in \cite{Visser}. If now we apply the constraint (\ref{7}), we can replace one of the constants $\xi$, {$k_3$,} $k_4$, $k_5$, $k_6$, or $k_7$.  For example, we can write
\ba
\xi &=& 20f_0-5k_3+\frac{5}{6}k_5+k_6+k_7 = 2^{11} \,3^2\,5^2 \,\pi \alpha -5k_3+\frac{5}{6}k_5+k_6+k_7\,,\n{8a}\\
k_6 &=& -\, 20f_0+\xi+5k_3-\frac{5}{6}k_5-k_7=- 2^{11} \,3^2\,5^2 \,\pi \alpha +\xi+5k_3-\frac{5}{6}k_5-k_7\,,\n{8}\\
k_7 &=& -\, 20f_0+\xi+5k_3-\frac{5}{6}k_5-k_6=- 2^{11} \,3^2\,5^2 \,\pi \alpha +\xi+5k_3-\frac{5}{6}k_5-k_6\, .\n{9}
\ea

Then, the equations for $\rho(z)$ and $P(z)$ can be written as follows:
\ba
\rho(z)&=&\frac{\beta z^2}{1-z}\biggl[\frac{\xi}{20}(1-20z^4+18z^5)+\frac{k_3}{4}(1-2z^2)+k_4z^2(1-z)\biggl.\nonumber\\
&&\biggl.\ \ \ \ \ \ \ \ \ \  -\,\frac{k_5}{24}(1-32z^3+30z^4)
-\,\frac{k_6}{20}(1-30z^4+28z^5)-\frac{k_7}{20}(1-32z^5+30z^6)\biggl]\nonumber\\
&=&\beta z^2\biggl[\frac{f_0}{1-z}(1-20z^4+18z^5)-\frac{k_3}{2}z^2(1+z-9z^2)+k_4z^2\biggl.\nonumber\\
&&\biggl.\ \ \ \ \ \ \ \ \ \ \ \ \ \ \ \ \ \ \ \ +\,\frac{k_5}{12}z^3(16-9z)
+\,\frac{k_6}{2}z^4-\frac{k_7}{2}z^4(2-3z)\biggl]\nonumber\\
&=&\beta z^2\biggl[\frac{f_{0}}{1-z}(1-30z^4+28z^5)+\frac{\xi}{2}z^4-\frac{k_3}{2}z^2(1+z-14z^2)+k_4z^2+\frac{k_5}{6}z^3(8-7z)-\frac{3k_7}{2}z^4(1-z)\biggl]\nonumber\\
&=&\beta z^2\biggl[\frac{f_0}{1-z}(1-32z^5+30z^6)-\frac{\xi}{2}z^4(2-3z)-\frac{k_3}{2}
z^2(1+z+z^2-15z^3)\biggl.\nonumber\\
&&\biggl.\ \ \ \ \ \ \ \ \ \ \ \ \ \ \ \ \ \ \ \ \ \ \ \ \ \ \ \ \ \ +\,k_4z^2
+\, \frac{k_5}{12}z^3(1+z)(16-15z)+\frac{3k_6}{2}z^4(1-z)
\biggl] \, ,\n{rhov2}\\
P(z)&=&\frac{\beta z^2}{1-z}\biggl[\frac{\xi}{20}(1-2z^5)+\frac{k_3}{4}(1-8z+6z^2)-k_4z^2(1-z)
\biggl.\nonumber\\
&&\biggl.\ \ \ \ \ \ \ \ \ \ -\,\frac{k_5}{24}(1+16z^3-18z^4)
-\, \frac{k_6}{20}(1+10z^4-12z^5)-\frac{k_7}{20}(1+8z^5-10z^6)\biggl]\nonumber\\
&=&\beta z^2\biggl[\frac{f_0}{1-z}(1-2z^5)-\frac{k_3}{2}z(4+z+z^2+z^3)-k_4z^2
\biggl.\nonumber\\
&&\biggl.\ \ \ \ \ \ \ \ \ \ \ \ \ \ \ \ \ \ \ \ \ \ \ \ \ \ \ \ \ \ \ -\,\frac{k_5}{12}z^3(8-z)
-\, \frac{k_6}{2}z^4-\frac{k_7}{2}z^5\biggl]\nonumber\\
&=&\beta z^2\biggl[\frac{f_{0}}{1-z}(1+10z^4-12z^5)-\frac{\xi}{2}z^4-\frac{k_3}{2}z(4+z+z^2+6z^3)-k_4z^2-\frac{k_5}{6}z^3(4-3z)+\frac{k_7}{2}z^4(1-z)\biggl] \nonumber\\
&=&\beta z^2\biggl[\frac{f_{0}}{1-z}(1+8z^5-10z^6)-\frac{\xi}{2}z^5-\frac{k_3}{2}z(4+z+z^2+z^3+5z^4)\biggl.\nonumber\\
&&\biggl.\ \ \ \ \ \ \ \ \ \ \ \ \ \ \ \ \ \ \ \ \ \ \ \ \ \ \ \ \ \ -\, k_4z^2-\frac{k_5}{12}z^3(8-z-5z^2)-\frac{k_6}{2}z^4(1-z)\biggl].\n{tauv2}
\ea

Expressions (\ref{rhov2}) and (\ref{tauv2}) are each presented in four forms, first in terms of $\xi$ and the $k_i$ coefficients, then in terms of $f_0$ and the $k_i$ coefficients and then by substituting either $k_6$ or $k_7$ from the constraint. This is done because current fits to the numerical computations of the energy-momentum tensor show that for spin-0 particles an expansion of (\ref{pform}) with terms up to $k_6$ are appropriate. However, for spin-1 particles, an expansion of (\ref{pform}) with terms up to $k_7$ are used in \cite{Bardeen:2017ypp}. 

\begin{center}
\begin{figure}[t]
\centering
 \includegraphics[width=8cm]{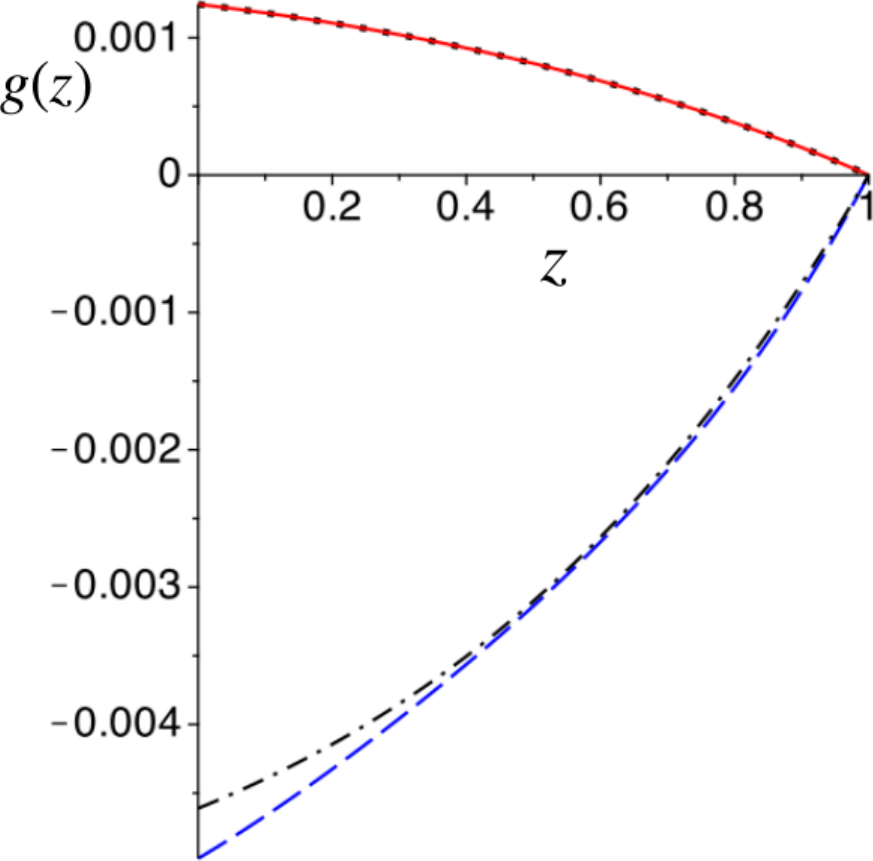}
   \caption{ Behaviour of the function $g(z)$ for spin 0, black dotted line for the case of nonzero $k_3$, $k_4$, $k_5$, and $k_6$ given by \cite{Bardeen:2017ypp} and red solid line for the case of nonzero $k_3$, $k_4$, $k_5$, and $k_6$ given by \cite{Bardeen:2018omt}. For spin 1, blue dashed line for the case of nonzero $k_3$, $k_4$, $k_5$, $k_6$, and $k_7$ given by \cite{Bardeen:2017ypp} and black dash-dotted line for the case of nonzero $k_3$, $k_4$, $k_5$, and $k_6$ given by \cite{Bardeen:2018omt}.\label{g}}
\end{figure}
\end{center}
\begin{center}
\begin{figure}[t]
\centering
\includegraphics[width=8cm]{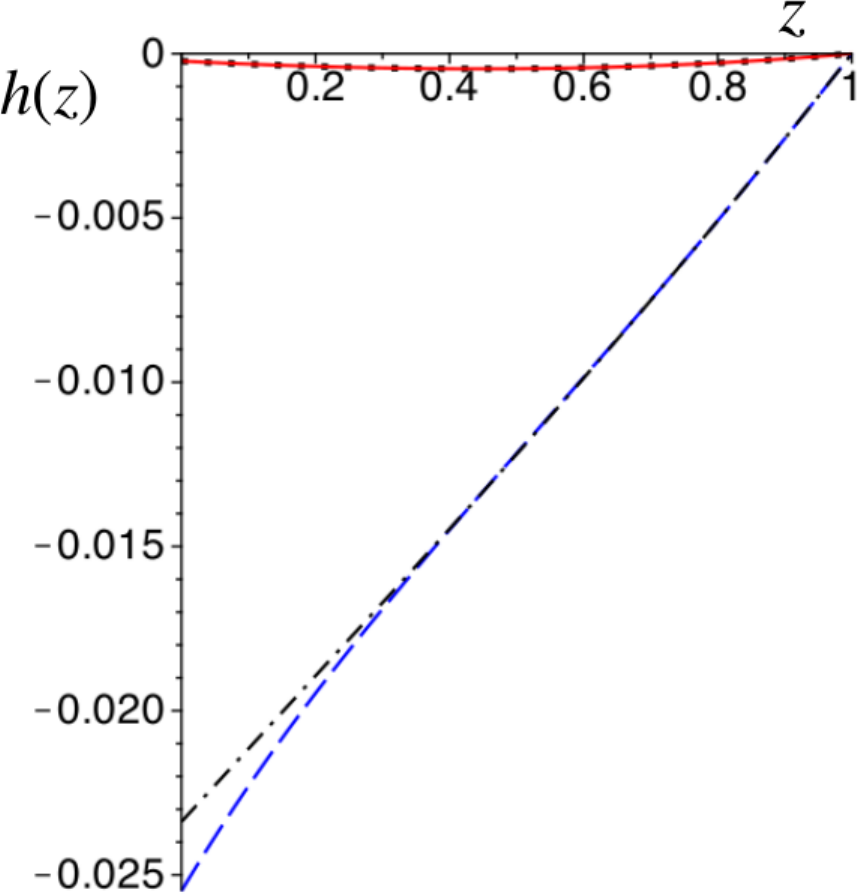}
   \caption{ Behaviour of the function $h(z)$ for spin 0, black dotted line for the case of nonzero $k_3$, $k_4$, $k_5$, and $k_6$ given by \cite{Bardeen:2017ypp} and red solid line for the case of nonzero $k_3$, $k_4$, $k_5$, and $k_6$ given by \cite{Bardeen:2018omt}. For spin 1, blue dashed line for the case of nonzero $k_3$, $k_4$, $k_5$, $k_6$, and $k_7$ given by \cite{Bardeen:2017ypp} and black dash-dotted line for the case of nonzero $k_3$, $k_4$, $k_5$, and $k_6$ given by \cite{Bardeen:2018omt}. \label{h}}
\end{figure}
\end{center}
\begin{center}
\begin{figure}[t]
\centering
  \includegraphics[width=8cm]{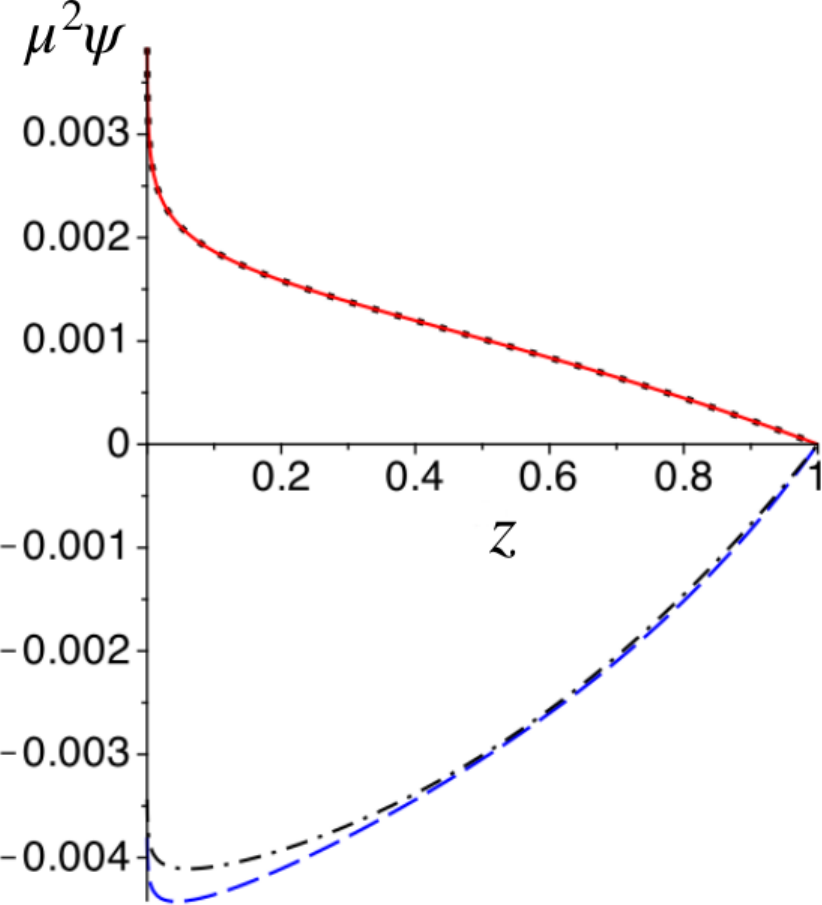}
   \caption{Behaviour of the function $\mu^2\psi$ for spin 0, black dotted line for the case of nonzero $k_3$, $k_4$, $k_5$, and $k_6$ given by \cite{Bardeen:2017ypp} and red solid line for the case of nonzero $k_3$, $k_4$, $k_5$, and $k_6$ given by \cite{Bardeen:2018omt}. For spin 1, blue dashed line for the case of nonzero $k_3$, $k_4$, $k_5$, $k_6$, and $k_7$ given by \cite{Bardeen:2017ypp} and black dash-dotted line for the case of nonzero $k_3$, $k_4$, $k_5$, and $k_6$ given by \cite{Bardeen:2018omt}. \label{B}}
\end{figure}
\end{center}

Now, solving the Einstein equation (\ref{EinsteinEq}), we obtain
\ba 
g(z)&=&\frac{1}{2^{11}3^{3}5^2\pi}(1-z)\biggl[2(\xi-k_6)(29+17z+8z^2)-5k_5(7+3z)-k_7(73+49z+31z^2+15z^3)\biggl]\nonumber\\
&=&\frac{1}{2^{11}3^{4}5\pi}(1-z)\biggl[6(4f_0-k_3)(29+17z+8z^2)+8k_5 (1+z+z^2)-9k_7(1+z)(1+z^2)\biggl]\nonumber\\
&=&\frac{1}{2^{12}3^{4}5\pi}(1-z)\biggl[-18(\xi-k_6)(1+z)(1+z^2)+6(4f_0-k_3)(73+49z+31z^2+15z^3)
+\, k_5(31+31z+31z^2+15z^3)\biggl], \nonumber
\ea
\ba
h(z)&=&\frac{1}{2^{11}3^{3}5^2\pi}(1-z)\biggl[-6\xi(3+5z+6z^2)+120k_4+5k_5(13+15z)+2k_6(19+25z+28z^2)
+k_7(23+35z+41z^2+45z^3)\biggl]\nonumber\\
&=&\frac{1}{2^{11}3^{3}5\pi}(1-z)\biggl[-6(4f_0-k_3)(3+5z+6z^2)+24k_4 
+2k_5(5+5z-3z^2) + 4k_6(1+z+z^2)+k_7(1+z+z^2+9z^3)\biggl]\nonumber\\
&=& \frac{1}{2^{11}3^{4}5\pi}(1-z)\biggl[12\xi(1+z+z^2)-6(4f_0-k_3)(19+25z+28z^2)
\biggl.\nonumber\\
&&\biggl.\ \ \ \ \ \ \ \ \ \ \ \ \ \ \ \ \ \ \ \ \ \
+72k_4 +4k_5(5+5z-7z^2)-9k_7(1-z)(1+2z+3z^2)\biggl]\nonumber\\
&=&\frac{1}{2^{12}3^{4}5\pi}(1-z)\biggl[6\xi(1+z+z^2+9z^3)-6(4f_0-k_3)(23+35z+41z^2+45z^3)
\biggl.\nonumber\\
&&\biggl. \ \ \ \ \ \ \ \ \ \ \  \ \ \ \ \ \ \ \ \ \ \
+144k_4+k_5(55+55z-41z^2-45z^3)+18k_6(1-z)(1+2z+3z^2)\biggl]
.\n{Chg}
\ea

Figures (\ref{g})-(\ref{B}) graph $g(z)$, $h(z)$, and $\mu^2\psi$ [given by Eq.\ (\ref{psirelation})] for the numerical values of the $k_i$ coefficients.

We remind the reader that $\xi$ is the value of the trace anomaly coefficient from Eq.\ (\ref{Tracerelation}). 
The constants {$k_3$,} $k_4$, $k_5$, $k_6$, and $k_7$ appear in Eq.\ (\ref{pform}) for the transverse pressure.  The expressions for $g(z)$ and $h(z)$ are each presented in four forms, first in terms of $\xi$ and the $k_i$ coefficients, then in terms of $f_0$ and the $k_i$ coefficients, and then by substituting either $k_6$ or $k_7$ from the constraint (except that the second and third forms for $g(z)$ coincide, so that form is written only once). The constant $f_0$ is related to the Hawking luminosity by $L = - dm/dv \approx \alpha/\mu^2=16\pi \beta f_0/\mu^2$, with $\beta\equiv1/({2^{13} 3^2 5\pi^2})$. 

For the case of a conformal scalar field (spin 0) in the Unruh state, Matt Visser \cite{Visser} has performed a least-squares fit to the transverse pressure data of Jensen, McLaughlin, and Ottewill \cite{Jensen:1991ef}
and of Anderson, Hiscock, and Samuel \cite{Anderson:1993if}, giving the constants $k_4=26.5652$, $k_5=-59.0214$, $k_6=38.2068$ (to six digits, though not claimed accurate beyond $1\%$), and $f_{0}=5.349$ (close to Elster's value \cite{Elster:1983pk} of 5.385 that is probably more accurate). From the summary in \cite{Visser} of the data of \cite{Jensen:1991ef, Anderson:1993if} and of private communications to Visser from some of those authors, Bardeen discovered that an improved fit to that data and to the scalar luminosity results \cite{ Simkins, Elster:1983pk, Taylor:1998dk} could be made by including a nonzero $k_3$, which he kindly provided us in 2016 \cite{Bardeenprivate} and published in 2017 \cite{Bardeen:2017ypp}: $k_3=0.264$, $k_4=25.438$, $k_5=-57.460$, $k_6=37.503$, and $f_{0}=5.385$. In 2018 Bardeen \cite{Bardeen:2018omt} presented revised fits of $k_3 = 0.2524$, $k_4 = 25.5439$, $k_5 = -57.6663$, $k_6 = 37.6172$, and $f_0 = 5.385$ for conformally invariant scalars (spin 0).
 
\begin{center}
\begin{figure}[t]
\centering
 \includegraphics[width=7cm]{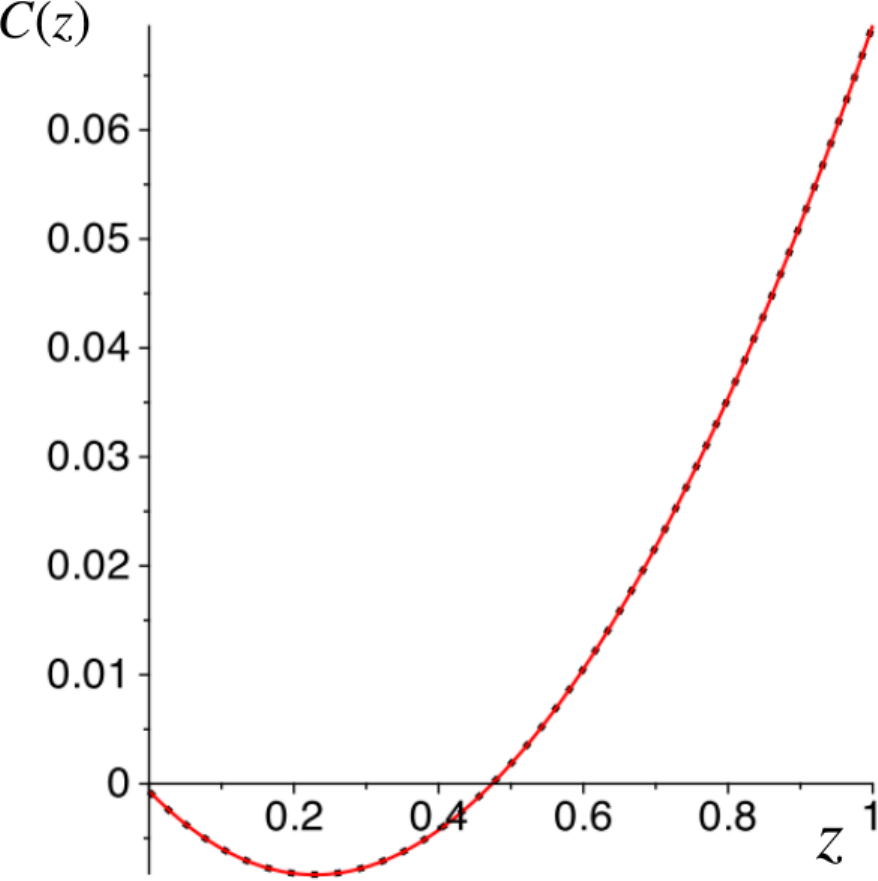}~~~~~~~
 \includegraphics[width=7cm]{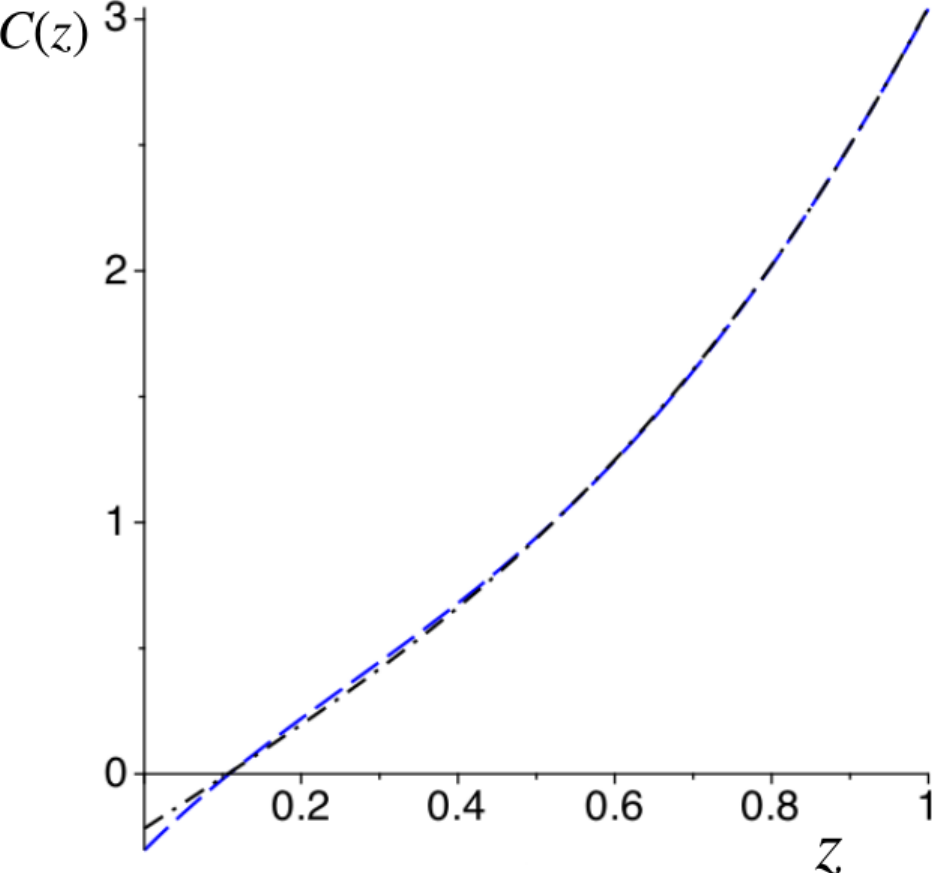}
   \caption{(a) Behaviour of the function $C \equiv r^6 R_{\alpha\beta\gamma\delta}R^{\alpha\beta\gamma\delta} - 48 m^2 \approx 2^{10} \pi z^{-3} [P(z) - \rho(z) - p(z)]$ for spin 0, black dotted line for the case of nonzero $k_3$, $k_4$, $k_5$, and $k_6$ given by \cite{Bardeen:2017ypp} and red solid line for the case of nonzero $k_3$, $k_4$, $k_5$, and $k_6$ given by \cite{Bardeen:2018omt}. (b) Behaviour of the function $C$ for spin 1, blue dashed line for the case of nonzero $k_3$, $k_4$, $k_5$, $k_6$, and $k_7$ given by \cite{Bardeen:2017ypp} and black dash-dotted line for the case of nonzero $k_3$, $k_4$, $k_5$, and $k_6$ given by \cite{Bardeen:2018omt}. \label{K2scalark3nzero}}
\end{figure}
\end{center}

\begin{center}
\begin{figure}[t]
\centering
  \includegraphics[width=7cm]{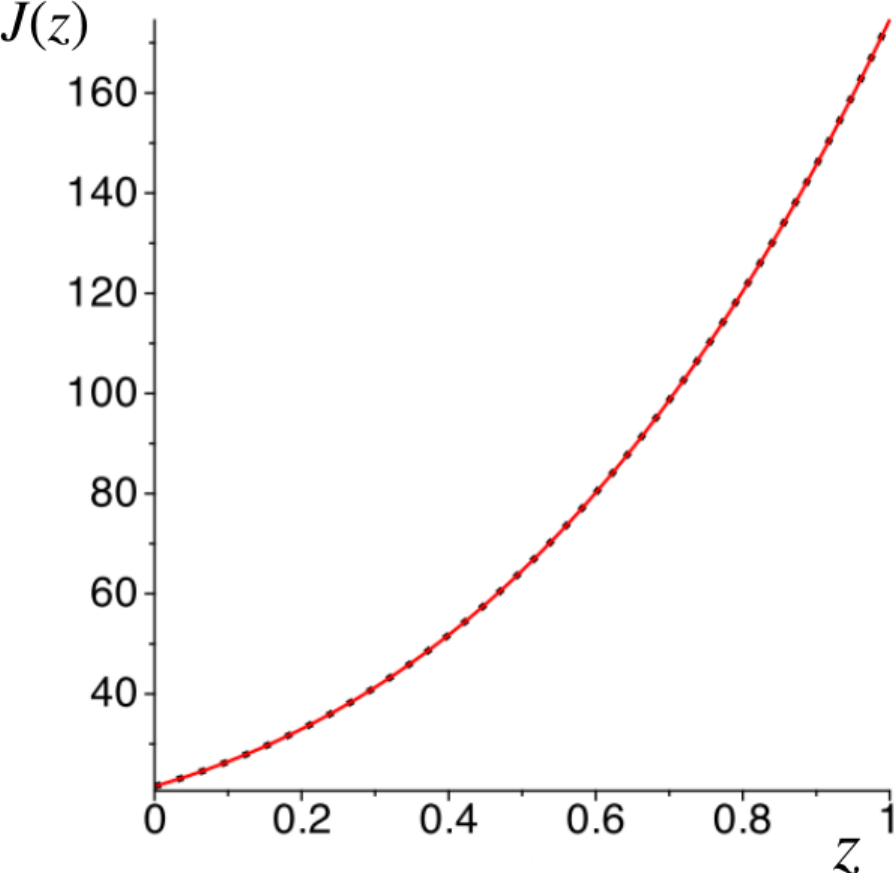}~~~~~~~
   \includegraphics[width=7cm]{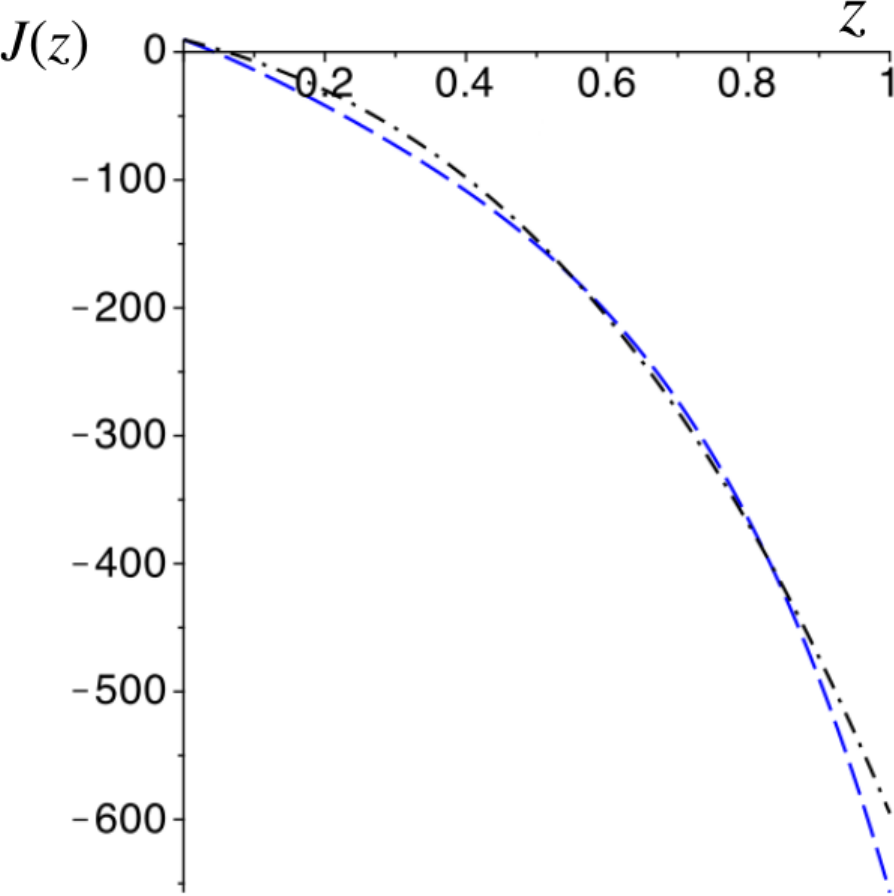}
   \caption{(a) Behaviour of the function $J(z)$ for spin 0, black dotted line for the case of nonzero $k_3$, $k_4$, $k_5$, and $k_6$ given by \cite{Bardeen:2017ypp} and red solid line for the case of nonzero $k_3$, $k_4$, $k_5$, and $k_6$ given by \cite{Bardeen:2018omt}. (b) Behaviour of the function $J(z)$ for spin 1, blue dashed line for the case of nonzero $k_3$, $k_4$, $k_5$, $k_6$, and $k_7$ given by \cite{Bardeen:2017ypp} and black dash-dotted line for the case of nonzero $k_3$, $k_4$, $k_5$, and $k_6$ given by \cite{Bardeen:2018omt}.
    \label{J}}
\end{figure}
\end{center}

\begin{center}
\begin{figure}[t]
\centering
  \includegraphics[width=7cm]{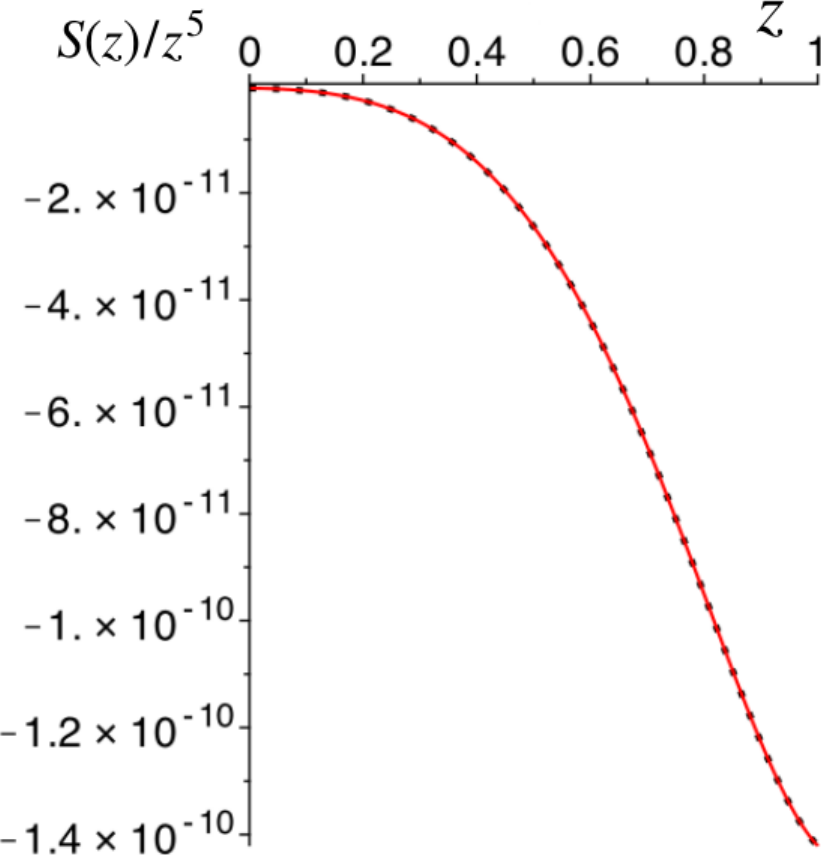}~~~~~~~
   \includegraphics[width=7cm]{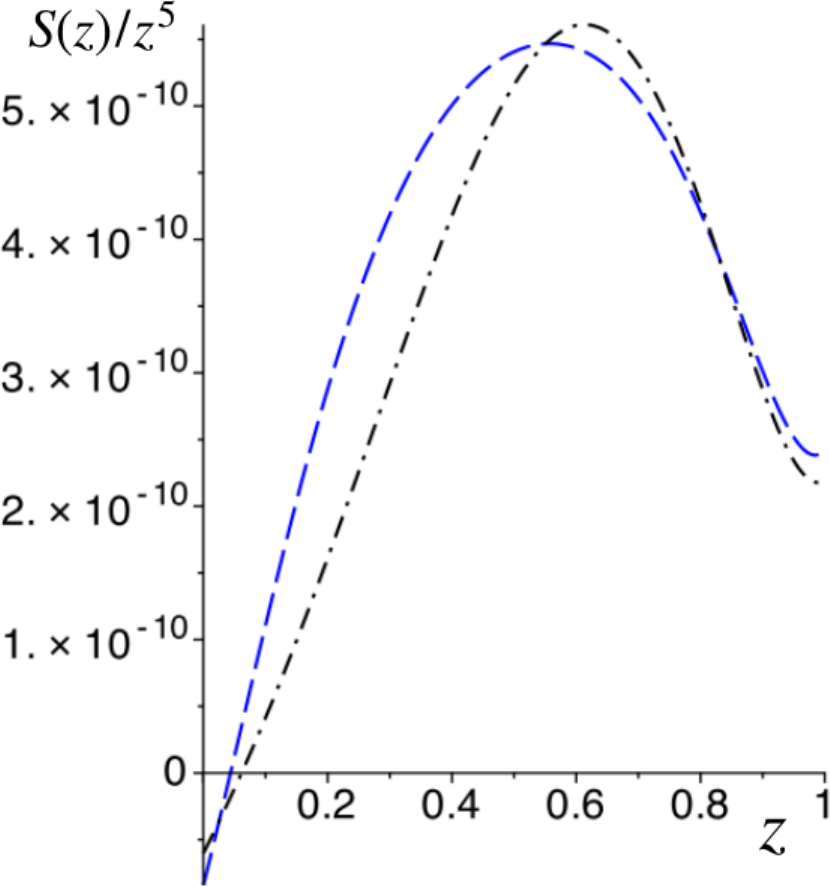}
   \caption{(a) Behaviour of the function $S(z)/z^5$ for spin 0, black dotted line for the case of nonzero $k_3$, $k_4$, $k_5$, and $k_6$ given by \cite{Bardeen:2017ypp} and red solid line for the case of nonzero $k_3$, $k_4$, $k_5$, and $k_6$ given by \cite{Bardeen:2018omt}. For spin 0, $S(z)<0$ for $0<z<1$, implying that the stress-energy tensor is of Hawking-Ellis Type IV \cite{HE} with no timelike or null eigenvector, so there is no frame without energy flux anywhere outside the black hole for the massless conformal scalar field for both cases of Bardeen's fitting coefficients and also for the earlier fit by Visser \cite{Visser} with $k_3 = 0$. (b) Behaviour of the function $S(z)/z^5$ for spin 1, blue dashed line for the case of nonzero $k_3$, $k_4$, $k_5$, $k_6$, and $k_7$ given by \cite{Bardeen:2017ypp} and black dash-dotted line for the case of nonzero $k_3$, $k_4$, $k_5$, and $k_6$ given by \cite{Bardeen:2018omt}. For $p(z)$ given by Eq.\ (\ref{pform}) for spin 1 with nonzero $k_3$, $k_4$, $k_5$, $k_6$, and $k_7$ given by \cite{Bardeen:2017ypp}, $S(z)>0$ for $z>0.04374$, but $S(z)<0$ for $0<z<0.04374$, where there is no frame with zero energy flux. For the spin-1 approximation for $p(z)$ with nonzero $k_3$, $k_4$, $k_5$, and $k_6$ given by \cite{Bardeen:2018omt}, $S(z)>0$ for $z>0.06151$, but $S(z)<0$ for $0<z<0.06151$.  However, if actually $k_3 = 0$, it is not yet clear whether or not there is any region with $S(z)<0$ for spin 1 (the electromagnetic field) that would give a Hawking-Ellis Type IV \cite{HE} stress-energy tensor with no timelike eigenvector.
   \label{S}}
\end{figure}
\end{center}

\begin{center}
\begin{figure}[t]
\centering
  \includegraphics[width=6cm]{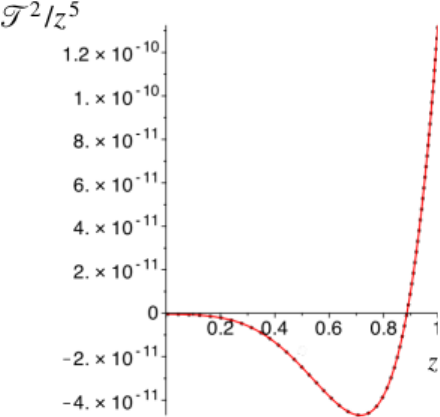}~~~~~~~
   \includegraphics[width=6cm]{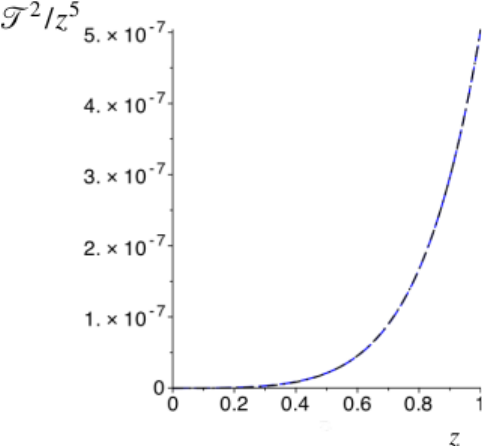}
   \caption{(a) Behaviour of the function $\mathcal{T}^2(z)/z^5$ for spin 0, black dotted line for the case of nonzero $k_3$, $k_4$, $k_5$, and $k_6$ given by \cite{Bardeen:2017ypp} and red solid line for the case of nonzero $k_3$, $k_4$, $k_5$, and $k_6$ given by \cite{Bardeen:2018omt}. For $p(z)$ given by Eq.\ (\ref{pform}) for spin 0 with nonzero $k_3$, $k_4$, $k_5$, $k_6$, and $k_7$ given by \cite{Bardeen:2017ypp}, $\mathcal{T}^2(z)>0$ for $z>0.88587$, but $\mathcal{T}^2(z)<0$ for $0<z<0.88587$, violating the TOSEC of \cite{Martin-Moruno:2013wfa}). For the spin-0 approximation for $p(z)$ with nonzero $k_3$, $k_4$, $k_5$, and $k_6$ given by \cite{Bardeen:2018omt}, $\mathcal{T}^2(z)>0$ for $z>0.88585$, but $\mathcal{T}^2(z)<0$ for $0<z<0.88585$. (b) Behaviour of the function $\mathcal{T}^2(z)/z^5$ for spin 1, blue dashed line for the case of nonzero $k_3$, $k_4$, $k_5$, $k_6$, and $k_7$ given by \cite{Bardeen:2017ypp} and black dash-dotted line for the case of nonzero $k_3$, $k_4$, $k_5$, and $k_6$ given by \cite{Bardeen:2018omt}.  For $p(z)$ given by Eq.\ (\ref{pform}) for spin 1 with nonzero $k_3$, $k_4$, $k_5$, $k_6$, and $k_7$ given by \cite{Bardeen:2017ypp}, $\mathcal{T}^2(z)>0$ for $z>0.01873$, but $\mathcal{T}^2(z)<0$ for $0<z<0.01873$. For the spin-1 approximation for $p(z)$ with nonzero $k_3$, $k_4$, $k_5$, and $k_6$ given by \cite{Bardeen:2018omt}, $\mathcal{T}^2(z)>0$ for $z>0.03039$, but $\mathcal{T}^2(z)<0$ for $0<z<0.03039$.  However, if actually $k_3 = 0$, it is not yet clear whether or not there is any region with $\mathcal{T}^2(z)<0$ for spin 1 (the electromagnetic field) that would violate the TOSEC of \cite{Martin-Moruno:2013wfa}).
\label{Tsquare}}
\end{figure}
\end{center}

For the electromagnetic field (massless spin 1) in the Unruh state, Bardeen \cite{Bardeen:2017ypp} used the results of Jensen and Ottewill \cite{Jensen:1988rh} and Jensen, McLaughlin, and Ottewill \cite{Jensen:1991ef} to give $k_3=114.62$, $k_4=-1186.24$, $k_5=1393.96$, $k_6=-2537.42$ and $k_7=652.20$, with $f_0=2.435$. In 2018 Bardeen \cite{Bardeen:2018omt} obtained access provided privately by Visser from the calculations of \cite{Jensen:1991ef}, and this led to an improved fit with $k_3=81.80$, $k_4=-770.42$, $k_5=65.38$, and $k_6=-942.18$, with $f_0=2.4346$ (using data from \cite{Page:1976ki}), with a nonzero $k_7$ no longer seen necessary.  For our Figs.\ (\ref{P})-(\ref{Tsquare}) we use the Bardeen \cite{Bardeen:2017ypp, Bardeen:2018omt} data as probably the most careful synthesis of previous calculations. 

Now that we have calculated the $h(z)$ and $g(z)$ functions of the metric (\ref{metric2}), we can calculate the Kretschmann scalar $\mathcal{K}\equiv R_{\alpha\beta\gamma\delta}R^{\alpha\beta\gamma\delta}$, which has the following form:
\ba
\mathcal{K} &=& 12F^2 + 4F(2G^a_a-G^i_i) + (G^a_a-G^i_i)^2 + 2 G^a_bG^b_a \nonumber\\
&=& 12F^2 + 8F(G^0_0+G^1_1-G^2_2) + 3(G^0_0)^2 + 3(G^1_1)^2 + 2G^0_0G^1_1 + 4G^0_1G^1_0 
    - 4G^2_2(G^0_0+G^1_1-G^2_2) \nonumber\\
&=& 12F^2 + 8F(G^r_r+G^v_v-G^\theta_\theta) + 3(G^r_r)^2 + 3(G^v_v)^2 +2G^r_rG^v_v
    + 4G^r_vG^v_r - 4G^\theta_\theta(G^r_r+G^v_v-G^\theta_\theta) \nonumber\\
&\approx& \frac{48m^2}{r^6}-\frac{128\pi m}{r^3\mu_0^4}\left[p(z)-P(z)+\rho(z)\right].
\ea 
Here, ${F}=2m/r^3$, and the indices $a$ and $b$ correspond to $0$ and $1$ while $i$ and $j$ correspond to $2$ and $3$. The first term in $\mathcal{K}$ is equal to the Kretschmann scalar for the Schwarzschild solution. In Fig.\ \ref{K2scalark3nzero} we have plotted 
\ba
C\equiv r^6\mathcal{K}-48m^2\approx \frac{2^{10}\pi}{z^3}\left[P(z)-\rho(z)-p(z)\right]= \frac{2^{10}\pi}{z^3} [T(z) - 3p(z)]\, .
\ea

We further define 
\be
S(z)=\mu_0^8(T^{ab}-\frac{1}{2}T^{c}_{~c}g^{ab})(T_{ab}-\frac{1}{2}T^{d}_{~d}g_{ab}),
\ee
where lowercase Latin letters $a,b$, etc.\ run only over $0$ and $1$, or $t$ and $r$.  In terms of

\be
J(z)=\frac{\rho+P+2f}{\beta~z^2(1-z)}=4f_0(1+2z+3z^2+4z^3)-k_3(2z+3z^2+4z^3)+\frac{2}{3}k_5z^3-k_7z^4~,
\ee this gives
\ba
S(z)&=& \frac{1}{2}(\rho+P)^2-2 f^2=\frac{1}{2}\beta^2{z^4 \, J(z)}\left[(1-z)^2J(z)-4f_0\right]\nonumber\\
&=&\frac{1}{18}\beta^2z^5\left[12f_0(1+2z+3z^2+4z^3)-3k_3(2z+3z^2+4z^3)+{2}k_5z^3-3k_7z^4\right]\nonumber\\
&&\times\left[-12f_0(5z^3-4z^4)-3k_3(2-z-5z^3+4z^4)+{2k_5}(z^2-2z^3+z^4)-3k_7(z^3-2z^4+z^5)\right] \nonumber\\
&=&\beta^2\left[-4f_0k_3 z^5+\left(-6f_0k_3+2k_3^2\right)z^6+\left(-8f_0k_3+\frac{4}{3}f_0k_5+2k_3^2\right)z^7   \right.
+\left(-40f_0^2-2f_0k_7+\frac{5}{2}k_3^2-\frac{4}{3}k_3k_5\right)z^8\nonumber\\
&&\ \ \ \ \ \  +\left(-48f_0^2+40f_0k_3-7{k_3^2}+\frac{2}{3}k_3k_5+2k_3{k_7}\right)z^9+\left(-56f_0^2+28f_0k_3-\frac{7}{2}k_3^2-k_3k_7+\frac{2}{9}k_5^2\right)z^{10}\nonumber\\
&&\ \ \ \ \ \ +\left(-64f_0^2+32f_0k_3-\frac{40}{3}f_0k_5-4k_3^2+\frac{10}{3}k_3k_5- \frac{4}{9}k_5^2-\frac{2}{3}k_5k_7\right)z^{11}\nonumber\\
&&\ \ \ \ \ \ +\left(128f_0^2-64f_0k_3+\frac{32}{3}f_0k_5+20f_0k_7+8k_3^2-\frac{8}{3}k_3k_5-5k_3k_7+\frac{2}{9}k_5^2+\frac{4}{3}k_5k_7+\frac{1}{2}k_7^2\right)z^{12}\nonumber\\
&&\left.\ \ \ \ \ \ +\left(-16f_0k_7+4k_3k_7-\frac{2}{3}k_5k_7-k_7^2\right)z^{13}+\frac{1}{2}k_7^2z^{14}\right].
\ea

Near the horizon, for $z\rightarrow 1$, we have
\be
S(z)\sim\beta^2f_0\left[-80f_0+18k_3-\frac{4}{3}k_5+2k_7 + \left(480f_0 -112k_3+\frac{28}{3}k_5-16k_7\right)\left(1-z\right)\right].
\ee
Figure (\ref{J}) shows the behavior of $J(z)$, and Fig.\ (\ref{S}) graphs $S(z)/z^5$.
 
\vspace{-1cm}


Note that very far from the black hole, $z \ll 1$, the leading contributions (lowest powers of $z$) are $S(z) \sim -4\beta^2 f_0 k_3 z^5 +\beta^2\left(-6f_0k_3+2k_3^2\right)z^6+\beta^2\left(-8k_3f_0+(4/3)f_0k_5+2{k_3^2}\right)z^7$.  Since for both spin 0 and spin 1, Bardeen's fits give $k_3 > 0$, if there is indeed such a positive $1/r^3$ term in the transverse stress, this would imply that $S(z)$ is negative for sufficiently small $z$ (large $r/(2\mu_0) \equiv 1/z$).  However, the $k_3 z^3$ term in Bardeen's fits to the transverse stress does not dominate over the $k_4 z^4$ term for either the conformal massless spin 0 or the massless spin 1 until one goes to values of $r$ much larger than the range where the stress could be calculated numerically, so although this term indeed improves the fit for the finite range of $r$ where the numerical calculations were made, possible small systematic errors in these calculations might have been responsible for the fit by a nonzero $k_3 z^3$ term, rather than indicating an actual asymptotic term of this form \cite{PrivateJames}.  Donald Marolf \cite{PrivateDonald} and Gary Horowitz \cite{PrivateGary} have expressed scepticism about the existence of a $1/r^3$ term in the transverse stress, and now Amos Ori \cite{PrivateAmos} has given us a detailed argument against such a term.  Therefore, if indeed asymptotically $k_3 = 0$, the leading term for $S(z)$ would be $(4/3) \beta^2 f_0 k_5z^7$, which is negative for conformal massless spin 0 but positive for massless spin 1.

Negative $S(z)$ means that the stress-energy tensor is of Hawking-Ellis Type IV \cite{HE}, so there is no timelike or null eigenvector.  This implies that no matter how fast an observer moves away from the black hole where $S(z) < 0$, the Hawking radiation energy flux in the observer's frame will always be outward (unlike solar radiation, for example, where at an average location of the earth at one astronomical unit from the sun, and using the value of the solar radius from \cite{Sun}, an observer with an outward velocity greater than $0.999\,993\,7453(12)$ of the speed of light [a gamma factor greater than $\gamma = 283.737(26) \approx 3^{1/4}$\! (astronomical unit)/(solar radius)] relative to the sun, will see the sun cover enough of sky [more than 63.4\%] that the flux will be inward in the observer's frame). 
Therefore, if indeed there were a nonzero $k_3 z^3$ term asymptotically for the transverse stress of the positive sign as given by Bardeen's fits \cite{Bardeen:2017ypp} and \cite{Bardeen:2018omt}, then for both the massless conformal spin 0 and spin 1, the stress-energy tensor would be asymptotically Type IV, with no frame in which the flux is zero, but if $k_3 = 0$, then the stress-energy tensor will be asymptotically Type IV for the conformal spin 0 with $k_5 < 0$ but the ordinary Type I (with a timelike eigenvector, for which an observer whose 4-velocity is the timelike eigenvector of the stress-energy tensor would see zero flux) asymptotically for massless spin 1 with $k_5 > 0$.

Numerically we find that the stress-energy tensor for the massless conformal scalar is Type IV everywhere outside the horizon for both of Bardeen's fits \cite{Bardeen:2017ypp} and \cite{Bardeen:2018omt} and also for the earlier fit by Visser \cite{Visser} with $k_3 = 0$. However, for massless spin 1, it is only Type IV for $0<z<0.04374$ for \cite{Bardeen:2017ypp} or for $0<z<0.06151$ for \cite{Bardeen:2018omt}, and since it would not be Type IV for sufficiently small $z$ if $k_3 = 0$ (assuming $k_5 > 0$ as given by both \cite{Bardeen:2017ypp} and \cite{Bardeen:2018omt}), we are not certain that the stress-energy tensor for the electromagnetic field in the Unruh state is Type IV (no timelike eigenvector) anywhere outside a slowly evaporating Schwarzschild black hole.

Here we should note that we disagree with the results \cite{Martin-Moruno:2013wfa}, who claim that the stress-energy tensor for a conformally coupled massless scalar field is Type I (the usual type, with a timelike eigenvector) near the horizon, for $z \gtrsim 0.9843$.  Their $\Gamma$, defined by their Eq.\ (2.28) to be the same as twice our $S$, is given by by a semianalytical model in their Eq.\ (7.41) (which seems to have many incorrect coefficients), and apparently these mistakes and/or round-off errors lead to the nonzero coefficient 0.9434 for the unphysical divergence at the horizon given by their Eq.\ (7.42).  We believe that we are the first to show that a conformally coupled massless scalar field in the Unruh state has a stress-energy tensor that is Hawking-Ellis Type IV \cite{HE} everywhere outside the horizon of a slowly evaporating Schwarzschild black hole, so that there are no observers anywhere outside that see zero energy flux.

Another scalar that we can calculate now that we know the form of the $h(z)$ and $g(z)$ functions is $\mathcal{T}^2\equiv \mu_0^8 T^{\mu \nu}T_{\mu \nu}$, which goes into the trace-of-square energy condition (TOSEC) of \cite{Martin-Moruno:2013wfa}, that $\mathcal{T}^2 \geq 0$:
\ba
\mathcal{T}^2(z) &=& \rho^2+P^2+2p^2-2f^2 = S + 2p^2 + \frac{1}{2}(2p - T)^2\, .
\ea
\ba
\mathcal{T}^2(z)&=&\beta^2\left[-4f_0k_3z^5+(-6f_0k_3+6k_3^2)z^6+\left(-8f_0 k_3+\frac{4}{3}f_0 k_5+2 k_3^2+8k_3k_4\right)z^7\right.\nonumber\\
&&\ \ \ \ \ \ +\left(-40f_0^2-2f_0k_7+\frac{5}{2}k_3^2+\frac{20}{3}k_3k_5+4k_4^2\right)z^8\nonumber\\
&&\ \ \ \ \ \ +\left(-48f_0^2+3k_3^2-k_3k_5+6k_3k_6+8k_4k_5\right)z^9\nonumber\\
&&\ \ \ \ \ \ +\left(-56f_{0}^2+28f_0k_3-40f_0k_4-\frac{7}{2}k_{3}^2+10k_3k_4+7k_3k_7-\frac{5}{3}k_4k_5+6k_4k_6-2k_4k_7+\frac{38}{9}k_5^2\right)z^{10}\nonumber\\
&& \ \ \ \ \ \ +\left(-64f_0^2+32f_0k_3-\frac{160}{3}f_0k_5-4k_3^2+\frac{40}{3}k_3k_5+8k_4k_7-\frac{19}{9}k_5^2+6k_5k_6-\frac{8}{3}k_5k_7\right)z^{11}\nonumber\\
&&\ \ \ \ \ \ +\biggl(328f_{0}^2-164f_0k_3+\frac{82}{3}f_0k_5-20f_0k_6+40f_0k_7 \nonumber\\
&&\ \ \ \ \ \ \ \ \ \ \ +\frac{41}{2}k_3^2-\frac{41}{6}k_3k_5+5k_3k_6-10k_3k_7
+\frac{41}{72}k_5^2 -\frac{5}{6}k_5k_6 +\frac{61}{6}k_5k_7+\frac{5}{2}k_6^2-k_6k_7+k_7^2\biggl)z^{12}\nonumber\\
&&\left.\ \ \ \ \ \ +\left(-56f_0k_7+14k_3k_7-\frac{7}{3}k_5k_7+6k_6k_7-3k_7^2\right)z^{13}+\frac{9}{2}k_7^2z^{14}\right]\, .
\ea

Near the horizon, for $z\rightarrow 1$, we have
\ba
\mathcal{T}^2(z)&\sim&\beta^2\left(120f_0^2-122f_0k_3-40f_0k_4-\frac{74}{3}f_0k_5-20f_0k_6-18f_0k_7+\frac{53}{2}k_3^2+18k_3k_4+\frac{73}{6}k_3k_5+11k_3k_6\right.\nonumber\\
&&\left.\ \ \ \ \ \ +11k_3k_7+4k_4^2+\frac{19}{3}k_4k_5+6k_4k_6+6k_4k_7+\frac{193}{72}k_5^2+\frac{31}{6}k_5k_6+\frac{31}{6}k_5k_7+\frac{5}{2}k_6^2+5k_6k_7+\frac{5}{2}k_7^2\right) \nonumber\\
&-&\beta^2\left(1920f_0^2-1448f_0k_3-400f_0k_4-\frac{748}{3}f_0k_5-240f_0k_6-264f_0k_7+264k_3^2\right.\nonumber\\
&&\left.\ \ \ \ \ \ +156k_3k_4+109k_3k_5+114k_3k_6+132k_3k_7+32k_4^2+\frac{166}{3}k_4k_5+60k_4k_6+68k_4k_7\right.\nonumber\\
&&\left.\ \ \ \ \ \ \ \ \ \ \ \ \ \ \ \ \ \ \ \ \ \ \ \ \ \ \ \ \ \ \ \ \ \ \ \ \ \ \ \ \ \  +\frac{155}{6}k_5^2+56k_5k_6+\frac{187}{3}k_5k_7+30k_6^2+66k_6k_7+36k_7^2\right)(1-z)
\ea

Numerically, we have that for the conformal massless scalar field, $\mathcal{T}^2(z) < 0$ (violating the TOSEC defined by \cite{Martin-Moruno:2013wfa}) for $z < 0.88587$ and $\mathcal{T}^2(z) > 0$ (the more common situation, obeying the TOSEC) for $0.88587 < z < 1$ from the fitting data of \cite{Bardeen:2017ypp}, or $\mathcal{T}^2(z) < 0$ for $z < 0.88585$ and $\mathcal{T}^2(z) > 0$ for $0.88585 < z < 1$ from the fitting data of \cite{Bardeen:2018omt}.  For the massless vector field, $\mathcal{T}^2(z) < 0$ for $z < 0.01873$ and $\mathcal{T}^2(z) > 0$ for $0.01873 < z < 1$ from the fitting data of \cite{Bardeen:2017ypp}, or $\mathcal{T}^2(z) < 0$ for $z < 0.03039$ and $\mathcal{T}^2(z) > 0$ for $0.03039 < z < 1$ from the fitting data of \cite{Bardeen:2018omt}.  However, these results assume the positive values of $k_3$ from Bardeen's fits \cite{Bardeen:2017ypp} and \cite{Bardeen:2018omt}), so that although with the more robust result that $k_5 < 0$ for the conformal massless scalar, the TOSEC for it seems to be violated at large distances from the black hole even with $k_3 = 0$, for the massless vector field (e.g., the electromagnetic field) with $k_5 > 0$, it is less clear whether or not the TOSEC is violated anywhere outside an evaporating black hole.

Figure (\ref{Tsquare}) shows the behavior of $\mathcal{T}^2(z)/z^5$. 

A lacuna in our numerical results is that there is as yet, so far as we know, no good approximations for the effective stress-energy tensor components from the spin-2 quantum gravitational field (other than for the flux coefficient $f_0$).  Assuming that photons and gravitons are the only massless fields in nature and that the lightest neutrino mass is not orders of magnitude smaller than the others so that it would be comparable to or less than the Hawking temperature of an astrophysical black hole, once we know the spin-2 stress-energy tensor of the Unruh state around a Schwarzschild black hole, then after the universe expands sufficiently, and after matter clears out from near each astrophysical black hole, then our formulas would enable one to calculate an accurate metric (that is, one whose small departure at each time from the Schwarzschild metric is accurately known) for such an isolated astrophysical black hole as it evaporates by the emission of Hawking radiation that would be almost entirely photons and gravitons.

So far, we have found the metric for $1 \ll (-3\alpha v)^{\frac{1}{3}} \lesssim r \ll -3v/2$.  (Indeed, it should apply even a bit inside the event horizon that is very near $z=1$, but for $z$ significantly larger than 1, the approximations used for the stress-energy tensor will not be valid.)  However, we have assumed that $2r \ll -3v$ so that the difference between $\mu(v,z)$ and $\mu_0(v) \equiv (-3\alpha v)^\frac{1}{3}$ is small, and so that $\tilde{z}$ in Eq.\ (\ref{psirelation}) is approximately $z \equiv 2\mu/r \approx 2\mu_0/r$.  Now it is time to find what $\tilde{z}$ is when these approximations are not valid, in order that we may have expressions for the metric coefficients that are good approximations no matter how large $r$ is for fixed $v$ and hence no matter how small $z$ is.  (Note that if $-4\alpha\ln{\tilde{z}}$ were replaced by $-4\alpha\ln{z}$ in Eq.\ (\ref{psirelation}), then $\psi$ would diverge as $r$ is taken to infinity and $z$ to zero, whereas actually $\psi$ stays finite and small.)

In a coordinate basis using the ingoing Eddington-Finkelstein coordinates $(v,r,\theta,\phi)$ of the metric (\ref{an1}), the Einstein equation gives
\be
\psi_{,r} = 4\pi r T_{rr} \, . \n{psir1}
\ee
Alternatively, in an orthonormal frame in which $e_{\hat{0}}$ is the 4-velocity of worldlines of constant $(r,\theta,\phi)$ and $e_{\hat{1}}$ is the unit spacelike vector in the outward radial direction orthogonal to $e_{\hat{0}}$, one has
\be
\psi_{,r} = \frac{4\pi r}{1-2m/r}
\left(T^{\hat{0}\hat{0}} +T^{\hat{0}\hat{1}} +T^{\hat{1}\hat{0}}+T^{\hat{1}\hat{1}}\right) \, . \n{psir2}
\ee
For the polynomial approximations for the stress-energy tensor given above, one gets that the radial partial derivative of $\psi$ at fixed ingoing null coordinate $v$, $\psi_{,r}$, is approximately $1/(\mu^2 r)$ multiplied by a polynomial in $z = 2\mu/r$. Note that we have made the gauge choice of setting $\psi(v,z=1)=0$ at the apparent horizon ($z \equiv 2\mu/r =1$), so that the value of $\psi(v,z)$ outside the apparent horizon (i.e., for $z < 1$) is obtained by integrating $\psi_{,r}dr$ from the apparent horizon at $r=2\mu$ to the greater value $r=2\mu/z$ along a null geodesic with constant $v$.  Because the integrand is approximately $1/(\mu^2 r)$ multiplied by a polynomial in $z$, we can integrate $1/(\mu^2 r)$ multiplied by each power of $z$ and then combine the results with the appropriate coefficients from the polynomial in $z$.  When we do this for the positive powers of $z$, $\mu$ will stay very near $\mu_0\equiv (-3\alpha v)^{\frac{1}{3}} \gg 1$ for nearly all of the dominant part of the integral, so that those integrals will combine to give $g(z)/\mu_0^2$. However, if we assume that $\mu \approx \mu_0$ when $1/(\mu^2 r)$ is multiplied by the constant term in the polynomial in $z$, this term integrates to $(-4\alpha/{\mu_0}^2)\ln{z}$, which diverges as $r$ is taken to infinity and hence $z$ is taken to zero.  This divergent logarithmic integral is of course dominated by very large $r$, where it is not valid to retain the approximation $\mu \approx \mu_0$, which is actually only valid for $2r \ll -3v$.  For larger $r$, $\mu$ rises with $r$, so that the integral of $dr/(\mu^2 r)$ remains bounded by a finite quantity no matter how large the upper limit of $r$ is taken.  What we actually get is the convergent integral
\be
(-4\alpha/{\mu_0}^2)\ln{\tilde{z}} = \int_{2\mu_0}^r \frac{4\alpha dr'}{\mu^2 r'}
\, . \n{lntildez}
\ee

This integral will give $\tilde{z} \approx z = 2\mu/r$ for $2r \ll -3v$.  The deviations will only become significant for larger $r$, for which the $1/z$-term on the right-hand side of Eq.\ (\ref{cubiceq}) becomes large in comparison with each of the other two terms.  Then we can let
\be
x \equiv \frac{\mu}{\mu_0} \approx 
  \left(1 + \frac{6\alpha r}{{\mu_0}^3}\right)^\frac{1}{3}=\left(1+\frac{2r}{-v}\right)^{\frac{1}{3}} \n{x}
\ee
become the new independent variable, which has a lower limit at $r = 2\mu_0$ of $x_h = (1+12\alpha/\mu_0^2)^\frac{1}{3} \approx 1 + 4\alpha/\mu_0^2$, which then gives
\be
-\ln{\tilde{z}} = \int_{x_h}^x \frac{3dy}{y^3-1} \, ,
\ee
and hence
\be
\tilde{z} \approx \frac{4\alpha}{\mu_0^2(x-1)}\sqrt{\frac{x^2+x+1}{3}}
\exp
\left[\sqrt{3}\tan^{-1}{\left(\frac{1}{\sqrt{3}}\frac{x-1}{x+1}\right)}\right] \, . \n{tildez}
\ee
For $6\alpha r \ll \mu_0^3 \equiv -3\alpha v$, so that $x \approx 1 + 2\alpha r/\mu_0^3$ is very near 1, this gives $\tilde{z} \approx z$, but unlike $z$, which goes to zero as $r$ goes to infinity, $\tilde{z}$ decreases only to the very small positive $v$-dependent constant
\be
\tilde{z}_0(v) \equiv \tilde{z}(v,z=0) \approx 
\frac{4\alpha}{\sqrt{3}\mu_0^2}\exp{\left(\frac{\sqrt{3}\pi}{6}\right)}
= \frac{4\alpha^{1/3}}{3^{7/6}(-v)^{2/3}}\exp{\left(\frac{\sqrt{3}\pi}{6}\right)} \, .
\ee 

As a result, the function $\psi$ in the ingoing Eddington-Finkelstein metric (\ref{an1}), which we have set to zero [along with $g(z)$ and $h(z)$] at the apparent horizon at $z=1$, does not diverge as $r$ is taken to infinity and $z \equiv 2\mu/r$ is taken to zero, but rather it goes to the finite (and small) limit of
\be
\psi_0(v) \equiv \psi(v,z=0) \approx \frac{(g_0 - 4\alpha \ln{\tilde{z}_0})}{\mu_0^2} \, .
\ee

In Figs.\ (\ref{g})-(\ref{B}), we have plotted the functions $g(z)$, $h(z)$, and $\mu^2\psi$ for spin-0 and spin-1 particles, assuming that $\mu_0$ is sufficiently large that the last plot does not extend into the very large $r \gtrsim -v$ regime where $\mu$ significantly exceeds $\mu_0$.

Therefore, knowing the value of the trace anomaly coefficients $\xi$ for the appropriate conformal massless fields, i.e., scalar and/or electromagnetic, and the constants {$k_3$,} $k_4$, $k_5$, $k_6$, and $k_7$, we have an approximate time-dependent metric for an evaporating black hole as a first-order perturbation of the Schwarzschild metric given by Eqs.\ (\ref{an1})-(\ref{m}) in $(v,r)$ coordinates with $z\equiv 2\mu/r$ from Eq.\ (\ref{z}), $\mu(v,z)$ from Eq.\ (\ref{mu_0})-(\ref{mu-cubic}) or $\mu(v,r)$ from Eq.\ (\ref{muapprox}) (approximately for $r \ll \mu^3/\alpha$), the functions $g(z)$ and $h(z)$ given by Eqs.\ ($\ref{Chg}$), and the function $\tilde{z}$ given by Eq.\ (\ref{tildez}) with $x$ therein given by Eq.\ (\ref{x}). 
We also remind the reader that we have chosen the gauge such that at the apparent horizon, where $z=1$ in our approximate metric, $\epsilon(1)=g(1)=h(1)=\ln{\tilde{z}(1)}=0$, and that this metric applies everywhere outside the black hole before it gets so small that quantum gravity effects become important.

\newpage
\section{Comparison of Metrics}
The metric ($\ref{an1}$) with our expressions for the functions contained therein should be valid for $1 \ll (-3\alpha v)^{\frac{1}{3}} \lesssim r$.  (Indeed, it should apply also for somewhat smaller $r$, a bit inside the event horizon that is very near $z=1$, but for $z$ significantly larger than 1, the approximations used for the stress-energy tensor will not be valid.)  Next, we want to convert it back to outgoing Eddington-Finkelstein coordinates to see how our metric behaves in the coordinate system $(u,r)$ in the restricted region $\mu/r=(-3\alpha u)^{1/3}/r\ll1$, where the outgoing Vaidya metric (\ref{Vadyaur}) is approximately valid. Consider the metric (\ref{an1}) with functions $h$ and $g$ known from Eq.\ (\ref{Chg}), using the relation
\be
v \approx u+2r-4 \left(-3\alpha u\right)^{1/3}+4\left(-3\alpha u\right)^{1/3}\ln\left[\frac{r}{2\left(-3\alpha u\right)^{1/3}}\right] \, .
\ee 
We get
\be
dv=\Psi_1du+\Psi_2dr \, ,
\ee
where
\ba
\Psi_1&\approx&1-\frac{4\alpha\ln\left[\frac{r}{2(-3\alpha u)^{1/3}}\right]}{(-3\alpha u)^{2/3}}+\frac{8\alpha}{(-3\alpha u)^{2/3}} \, , \\
\Psi_2&\approx&2+\frac{4(-3\alpha u)^{1/3}}{r} \, .
\ea 
The metric (\ref{an1}) with functions $h$ and $g$ given by Eq.\ (\ref{Chg}) becomes
\ba
&&ds^2=-e^{2\psi}\left(1-\frac{2m}{r}\right)\Psi_{1}^{2}du^2-2\left[e^{2\psi}\left(1-\frac{2m}{r}\right)\Psi_{1}\Psi_{2}-e^{\psi}\Psi_{1}\right] du dr \nonumber\\
&&\ \ \ \ \ \ \ \ \ +\left[2e^{\psi}\Psi_{2}-e^{2\psi}\left(1-\frac{2m}{r}\right)\Psi_2^2\right]dr^2+r^2 d\Omega^2 \nonumber\\
&&\ \ \ \ \ =g_{uu}du^2+2g_{ur} du dr+g_{rr}dr^2+r^2 d\Omega^2 \, . \n{ur}
\ea
For $-3\alpha u\gg 1$ we have
\ba
&&g_{uu}=-1+\frac{2(-3\alpha u)^{1/3}}{r}+\frac{1}{(-3\alpha u)^{1/3} \, r}\left[32\alpha+2h+4g\right]+\mathcal{O}(\frac{1}{(-3\alpha u)^{2/3}}) \, ,
\n{guumetric} \\
&&g_{ur}=-1+\frac{8(-3\alpha u)^{2/3}}{r^2}+\mathcal{O}(\frac{1}{r^2}) \, ,\n{gurmetric} \\
&&g_{rr}=\frac{16(-3\alpha u)^{2/3}}{r^2}+\frac{-96\alpha u}{r^3}+\mathcal{O}(\frac{(-3\alpha u)^{4/3}}{r^4}) \,\n{grrmetric} .
\ea 
Equations (\ref{ur})-(\ref{grrmetric}) give us the corrections to the outgoing Vaidya metric (\ref{Vadyaur}) for an evaporating black hole in outgoing Eddington-Finkelstein coordinates.  
When we consider only terms of the order of unity and of first order in the small quantity $\mu/r=(-3\alpha u)^{1/3}/r$, we get
\ba
&&g_{uu} \approx -1+\frac{2(-3\alpha u)^{1/3}}{r} \, , \\
&&g_{ur} \approx -1 \, , \\
&&g_{rr} \approx 0 \, .
\ea 
Therefore, for $\mu/r=(-3\alpha u)^{1/3}/r \ll 1$ we get the outgoing Vaidya metric,
\ba
&&ds^2\approx\left(-1+\frac{2(-3\alpha u)^{1/3}}{r}\right)du^2-2 du dr+r^2 d\Omega^2 \, . \n{ur2}
\ea

This form of the metric applies for $1 \ll (-3\alpha v)^{\frac{1}{3}} \ll r$, so it does not apply near the black hole horizon, where the more general Eq.\ ($\ref{an1}$) does apply, but it is applicable for arbitrarily large $r$.  It also applies for positive advanced time $v$, after the black hole has evaporated, so long as one avoids the Planckian region near the final evaporation and its causal future, where quantum gravity effects are expected.

\section{Summary}
In this paper, we have constructed an approximate time-dependent metric for an evaporating black hole as a first-order perturbation of the Schwarzschild metric, using the linearized backreaction from the stress-energy tensor of the Hawking radiation in the Unruh quantum state in the unperturbed spacetime. We used a metric ansatz in ingoing Eddington-Finkelstein coordinates $(v,r)$. Our ansatz is such that at infinity we get the Vaidya metric in the outgoing Eddington-Finkelstein coordinates $(u,r)$. We have solved the corresponding Einstein equation for the metric functions to first order in the stress-energy tensor of the unperturbed Schwarzschild metric. Therefore, our metric should be a very good approximation everywhere near to and everywhere outside the event horizon when the mass is large in Planck units.

\section*{Acknowledgments}
We acknowledge helpful information about the stress-energy tensor by email from James Bardeen. This work was
supported in part by the Natural Sciences and Engineering Research Council of Canada. During the latter stages of the paper, DNP appreciated the hospitality of the Perimeter Institute and of Matthew Kleban at the Center for Cosmology and Particle Physics of New York University. Research at Perimeter Institute is supported by the Government of Canada through the Department of Innovation, Science and Economic Development and by the Province of Ontario through the Ministry of Research, Innovation and Science.  Calculations by DNP during one stage of the paper were made at the Cook's Branch Nature Conservancy under the generous hospitality of the Mitchell family and of the George P. and Cynthia W. Mitchell Institute for Fundamental Physics and Astronomy of Texas A $\&$ M University. Private communications of SA with Donald Marolf 2019 August 7 were during SA's visit to the Kavli Institute for Theoretical Physics (KITP) at the University of California, Santa Barbara. SA used the KITP's facilities as a KITP scholar supported in part by the National Science Foundation under Grant No. NSF PHY-1748958.

\end{document}